\documentclass[twocolumn,superscriptaddress,pra,nofootinbib]{revtex4-1}
\usepackage{times}
\usepackage{color}
\usepackage{xspace}
\usepackage{amsmath}
\usepackage{amssymb}
\usepackage{amsbsy}
\usepackage{graphicx}
\usepackage{bm}
\usepackage{float}
\usepackage{relsize}
\usepackage[normalem]{ulem}

\usepackage[unicode,breaklinks]{hyperref}
\hypersetup{
    unicode=true,
    plainpages=false, 
    colorlinks=true,
    linkcolor=blue,
    citecolor=blue,
    filecolor=black,
    urlcolor=blue
}

\newcommand{\gdd}{g^\mathrm{dd}}
\newcommand{\edd}{\epsilon^\mathrm{dd}}
\newcommand{\bx}{\mathbf{x}}
\newcommand{\br}{\mathbf{r}}
\newcommand{\bk}{\mathbf{k}}
\newcommand{\mum}{\mu^m}
\newcommand{\Ut}{\tilde{U}}

\begin{document}

\title{Miscibility and stability of dipolar bosonic mixtures}

\author{Au-Chen Lee}
\affiliation{Department of Physics, Centre for Quantum Science,
and Dodd-Walls Centre for Photonic and Quantum Technologies, University of Otago, Dunedin, New Zealand}
\author{D. Baillie}
\affiliation{Department of Physics, Centre for Quantum Science,
and Dodd-Walls Centre for Photonic and Quantum Technologies, University of Otago, Dunedin, New Zealand}
\author{P. B. Blakie}
\affiliation{Department of Physics, Centre for Quantum Science,
and Dodd-Walls Centre for Photonic and Quantum Technologies, University of Otago, Dunedin, New Zealand}
\author{R. N. Bisset}
\affiliation{Institut f\"ur Experimentalphysik, Universit\"at Innsbruck, Innsbruck, Austria}

\begin{abstract}

Combining two Bose-Einstein condensates (BECs) may result in a miscible or immiscible mixture, or even a violent implosion.
We theoretically demonstrate that dipolar two-component BECs produce far richer physics than their nondipolar counterparts.
Intriguingly, when both components have equivalent dipoles, the transition to immiscibility is largely unaffected by dipolar physics, yet the dipoles maximally affect stability.
Conversely, antiparallel dipoles strongly affect miscibility but have little effect on stability.
By performing three-dimensional calculations of the ground states and their excitations, we find strong dependencies on the confinement geometry. We explore and elucidate the various phononic and rotonic phase transitions, as well as symmetry preserving crossovers.

\end{abstract}

\maketitle


\section{Introduction}

Two-component Bose-Einstein condensates (BECs) exhibit an intriguing phase diagram.
Starting from an immiscible (phase separated) ground state, and then decreasing the strength of the intercomponent contact interactions, the ground state transitions to a miscible (cospatial) state \cite{Hall1998B,Papp2008B,McCarron2011,Lercher2011a,Wacker2015,Wang2016,Ho1996,Esry1997,Pu1998,Timmermans1998}.
Further decreasing the intercomponent interactions, into the attractive regime, eventually results in a violent implosion,
even if both BECs would be individually stable.\footnote{Note that there is a narrow range of intercomponent contact interaction strengths in which beyond-meanfield quantum fluctuations may stabilize against such an implosion, resulting in a two-component self-bound droplet \cite{Petrov2015a,Cabrera2018,Semeghini2018}.}

Dipolar two-component BECs are expected to be far richer than their nondipolar counterparts, and the recent experimental realization of dual-species BECs consisting of highly-magnetic rare-earth elements---erbium and dysprosium---opens the door to numerous exciting possibilities \cite{Trautmann2018,Durastante2020}.
Single-component dipolar BECs have already proven to be incredibly interesting with the recent experimental realization of ultradilute droplets with liquid-like properties \cite{Kadau2016a,Chomaz2016,Schmitt2016a}, rotonic excitation spectra \cite{Chomaz2018a,Petter2019}, and the long-sought supersolid phase \cite{Tanzi2019,Bottcher2019,Chomaz2019}.

The pioneering theoretical work by G\'oral and Santos investigated the special case of miscible BECs with antiparallel dipoles---where the dipoles in one component are antiparallel to those of the other---in the absence of contact interactions \cite{Goral2002}. Although such BECs would be unstable in the homogeneous limit, they can be stabilized by quantum pressure (a kinetic energy effect) for small atom numbers. Intriguingly, when the components have balanced populations they predicted the ground state to simply be that of a noninteracting harmonic oscillator, right up to the instability threshold, after which the BECs would implode.

More recent theoretical studies in flattened trapping geometries predict that two-component (or binary) BECs, consisting of a dipolar and a nondipolar component, can undergo a rotonic miscible-immiscible transition \cite{Wilson2012a}, or produce an interface ferrofluid capable of exhibiting supersolidity \cite{Saito2009}.
Dipolar mixtures with antiparallel dipoles have also been predicted to display finger instabilities in the immiscible phase \cite{KuiTian2018},
while rotating dipolar mixtures may exhibit exotic vortex lattices \cite{Zhang2015,Zhang2016,Kumar2017}, including half-quantum vortex molecules \cite{Shirley2014}. The interplay of rotation and the miscible-immiscible transition has been theoretically studied by either adjusting the direction of dipole polarization \cite{Tomio2020} or by tuning the dipole-dipole interaction coefficient \cite{Kumar2019,Giovanazzi2002}, a technique that was recently demonstrated in experiments \cite{Tang2018}.
A number of other theoretical works have also focused on flattened geometries  \cite{Jain2011,Young-S2012,Kumar2017B}, as well as highly elongated, quasi-one dimensional configurations \cite{Gligoric2010,Hocine2019}.

Here we theoretically investigate dipolar binary BECs, performing a systematic study of the effects arising from the two components having differing dipole magnitudes and relative orientations, ranging in a continuous way from the parallel to the antiparallel situation.\footnote{Reference \cite{Goral2002} considered the possibility of antiparallel dipoles in the context of diatomic molecules. However, given the recent experimental advances with highly magnetic rare-earth elements, it may be more practical to consider different spin projections, where appropriate experimental methods can be implemented to suppress dipolar relaxation \cite{Chalopin2020}. We note that our results are applicable for both electric and magnetic dipolar mixtures.}
While parallel dipole moments maximally affect mechanical instability to implosion, which is driven by the in-phase fluctuations of the two densities, antiparallel dipoles instead strongly affect the miscible-immiscible threshold that is driven by the out-of-phase (or pseudo-spin) fluctuations of the two densities. Note that we herein refer to the latter simply as spin fluctuations.
By solving a three-dimensional (3D) Gross-Pitaevskii equation (GPE) we find a strong interplay between the dipolar effects and the underlying confinement geometries, which we take to be cylindrically symmetric with the dipoles aligned (or antialigned) along their symmetry axes.
To characterize the various phase transitions and crossovers, we calculate the excitations using a 3D Bogoliubov--de Gennes (BdG) theory.
We also calculate the dynamic structure factor---a quantity recently utilized in experiments to reveal the dipolar roton \cite{Petter2019}---to elucidate how phonon instabilities present in elongated (or cigar-shaped) confining traps can give way to roton instabilities for flattened  (or pancake-shaped) geometries.\\

In Sec.~\ref{Sec:Form} we describe the GPE formalism employed for our stationary state calculations, as well as the BdG theory for the corresponding excitation energies. Section \ref{Sec:Homo} discusses mechanical instability and immiscibility of two-component dipolar BECs in the homogeneous limit, without external trapping. Our main results are presented in Sec.~\ref{Sec:Trapped}, for which we consider the important effects of 3D confinement on the phase diagram. We also study excitations and the role they play for instability and immiscibility. Section \ref{Sec:SAIP} investigates the competing symmetric-immiscible and asymmetric-immiscible stationary states. Our conclusions are presented in Sec.~\ref{Sec:Conc}.

\section{Formalism}\label{Sec:Form}

Our stationary state solutions are obtained using a two-component dipolar GPE, while the corresponding excitations are calculated with a BdG theory.

\subsection{Meanfield theory}

To obtain the condensate wavefunction for each component $\psi_i$ (where $i=1,2$), which we take to be real, 
we solve the two-component dipolar GPE \cite{Goral2002},
\begin{equation}
\mathcal{L}_i\psi_i(\bx) = \mu_i\psi_i(\bx) ,
\end{equation}
where the GPE operator is given by
\begin{equation}
\mathcal{L}_i = -\frac{\hbar^2\nabla^2}{2M} + V(\bx) + \sum_{j=1}^2\int d\bx' U_{ij}(\bx-\bx') |\psi_j(\bx')|^2, \label{Eq:GPEOp}
\end{equation}
and $\mu_i$ is the chemical potential of component $i$. We consider harmonic trapping potentials $V(\bx)=\frac{1}{2}M[\omega_\rho^2(x^2+y^2)+\omega_z^2z^2]$,
where we take the frequencies $\omega_\rho$, $\omega_z$ and the mass $M$ to be the same for both components.\footnote{
For optical dipole traps, it is a good approximation to assume equal trapping frequencies for the various isotopic combinations of Dy and Er since they have comparable atomic masses and atomic polarizabilities \cite{Trautmann2018}.
Trapping frequencies can also be controlled for more general mixtures by tuning the atomic polarizability, which can be achieved by various experimental techniques such as adjusting the laser frequency \cite{Grimm2000}.}
The trap aspect ratio is $\lambda=\omega_z/\omega_\rho$, and the dipoles are aligned either parallel or antiparallel to the $z$ axis.
The indices $i$ and $j$ denote the interactions between (or within) components, and thus the two-body contact and dipole-dipole interactions (DDIs) take the form, respectively,
\begin{equation}
U_{ij}(\br) = g_{ij}\delta(\br) + \frac{3\gdd_{ij}}{4\pi} \frac{1-3\cos^2\theta}{r^3}.
\end{equation}
Here, $g_{ij} = 4\pi\hbar^2a_{ij}/M$ and $\gdd_{ij} = 4\pi\hbar^2a^{dd}_{ij}/M$, where $a_{ij}$ is the $s$-wave scattering length, $a^{dd}_{ij} = M\mu_0\mum_i\mum_j/12\pi\hbar^2$ is the DDI length, $\mum_i$ is the magnetic moment, and $\theta$ is the angle between the axis linking the two particles and the dipole polarization axis, $z$.

\subsection{Excitations}
To predict the dynamical stability of a state, we linearize the time-dependent GPE about the solution and express the expansion as
\begin{equation}
\Psi_j(\bx,t)=e^{-i\mu_j t/\hbar}\left[\psi_j(\bx)+\vartheta_j(\bx,t)\right],
\end{equation}
where
\begin{equation}
\vartheta_j(\bx,t)\equiv\sum_{\nu}\left(c_\nu u_{\nu,j} e^{-i\epsilon_\nu t/\hbar}-c_\nu^{*}v_{\nu,j}^{*}e^{i\epsilon_\nu^{*}t/\hbar}\right)
\end{equation}
is the fluctuation part, $c_\nu$ is the perturbation amplitude, while $u_\nu$ and $v_\nu$ are the quasiparticle amplitudes with excitation energy $\epsilon_\nu$. Dynamical instability is signaled by the presence of at least one $\epsilon_\nu$  being imaginary or complex.

The Bogoliubov--de Gennes equations can be written as a $4\times4$ BdG matrix $H\mathbf{u}=\epsilon\mathbf{u}$, where $\mathbf{u}= (u_{\nu,1}, u_{\nu,2}, v_{\nu,1}, v_{\nu,2})^T$ and
\begin{equation}
H=
\begin{pmatrix}
\mathbf{L}+\mathbf{X}&-\mathbf{X}\\
\mathbf{X}&-\mathbf{L}-\mathbf{X}
\end{pmatrix} ,
\end{equation}
which contains two submatrices
\begin{equation}
\mathbf{L}=\begin{pmatrix}
\mathcal{L}_{1}-\mu_1&0\\
0&\mathcal{L}_{2}-\mu_2
\end{pmatrix},
\end{equation}
where $\mathcal{L}_{i}$ is given by Eq.~\eqref{Eq:GPEOp}, and the components of the exchange operator $\mathbf{X}$ are
\begin{align}
    X_{ij}f(\bx)=\psi_i(\bx)\int d\bx' U_{ij}(\bx-\bx')\psi_j(\bx')f(\bx') .
\end{align}
\subsection{Structure Factor}
The density ($+$) and the spin ($-$) dynamic structure factors are
\begin{align}
S_\pm(\bk,\omega)&=\sum_{\nu}|\delta n^{\pm}_{\bk,\nu}|^2\delta(\omega-\omega_\nu) ,\label{Eq.SF}
\end{align}
where
\begin{align}
\delta n^{\pm}_{\bk,\nu}&\equiv\int d\bx e^{-i\bk\cdot\bx}(\delta n_{\nu,1}\pm\delta n_{\nu,2})
\end{align}
is the Fourier transform of the density and spin fluctuation, with
\begin{align}
\delta n_{\nu,j}&={\psi}_j({u}_{\nu,j}-{v}_{\nu,j})
\end{align}
being the density fluctuation of component $j$.

\subsection{Numerical methods}

We obtain stationary states of the GPE in a two-step process, first using a gradient flow method \cite{Bao2006} (also see Ref.~\cite{Lee2020}), and then a Newton-Krylov solver \cite{Kelley2003}. The latter offers improved convergence, but gives better performance with an initial state that is close to the desired stationary state.
We used the Krylov-Schur algorithm for the BdG excitations.
We take advantage of cylindrical symmetry by using Bessel grids and Fourier-Hankel transforms \cite{Ronen2006a}.
Furthermore, we employ the cylindrical cutoff interaction potential derived in Ref.~\cite{Lu2010a} to calculate the DDIs.

\begin{figure}
\begin{center}
\includegraphics[width=3.4in]{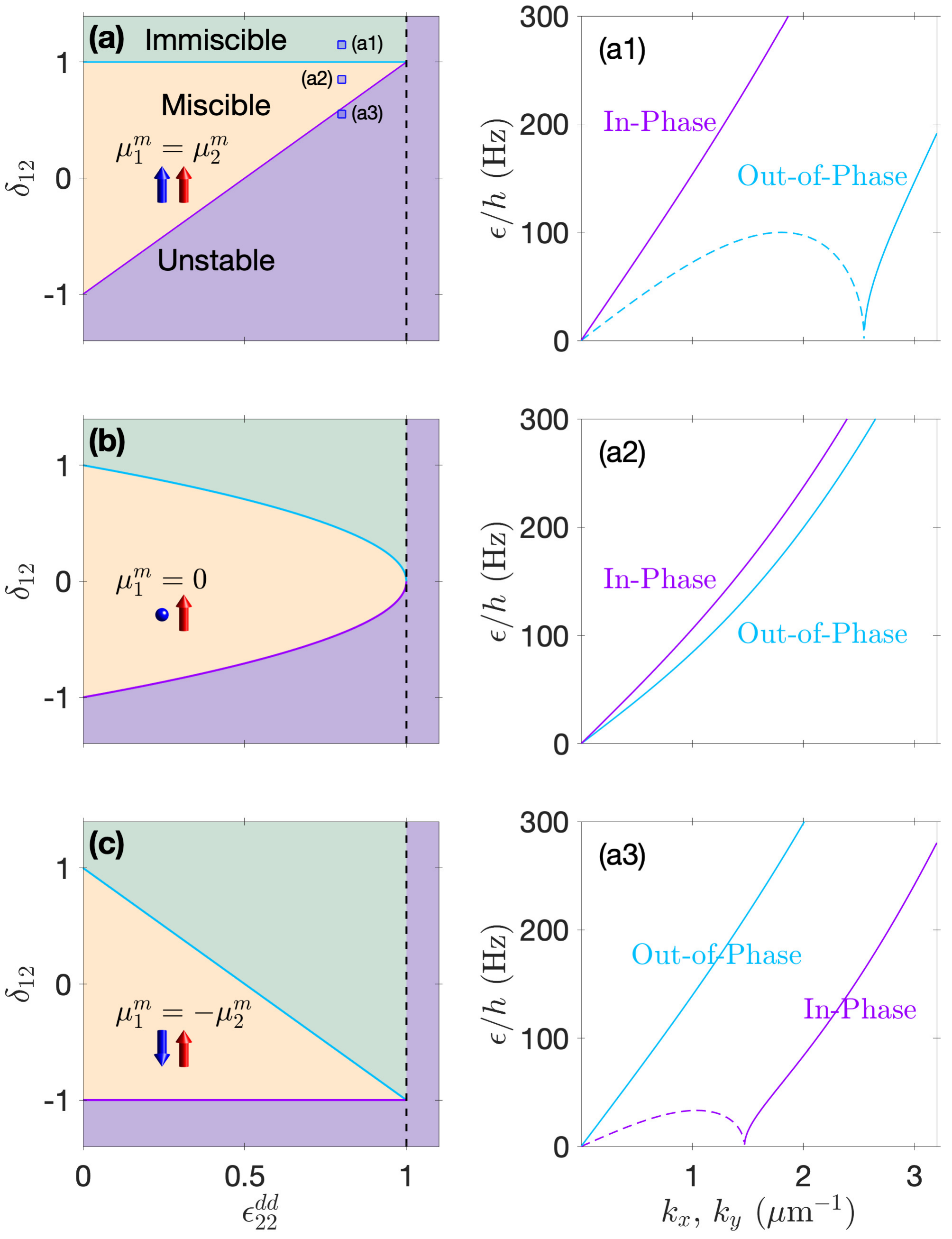}
\caption{Homogeneous phase diagrams [based on Eq.~(\ref{Eq:ho_ineq_edd})] for two-component dipolar BECs showing immiscible (green), miscible (orange), and mechanically unstable (purple) phases.
The $x, y$ axes represent the relative dipole strength [Eq.~(\ref{Eq:rel_edd})] and relative interspecies scattering length [Eq.~(\ref{Eq:rel_a12})], respectively.
The three cases are for (a) two components with aligned dipoles, (b) a dipolar with a nondipolar component, and (c) two components with antiparallel dipoles.
The unstable regions to the right of the vertical dashed lines indicate $\epsilon^{dd}_{ii}>1$, where the dipolar components are individually unstable.
(a1--a3) Excitation spectra for the density (in-phase) and spin (out-of-phase) branches of Eq.~(\ref{homogEpm}), with parameters marked by the three blue squares in (a). Parameters: $\theta_k=\pi/2,~n_1=n_2=10^{20}\:\mathrm{m}^{-3},~M=164\,\mathrm{u},~a^{dd}_{11}=a^{dd}_{22}=130a_0$ [but $a^{dd}_{11}=0$ for (b)], $a_{11}=a_{22}=162.5a_0$, (a1) $\delta_{12}=1.15$, (a2) $\delta_{12}=0.85$ and (a3) $\delta_{12}=0.55$.}\label{Fig:figure1}
\end{center}
\end{figure}

\section{Homogeneous mixtures}\label{Sec:Homo}

Basic understanding of mechanical instability and immiscibility of a miscible binary BEC can be developed from a simple homogeneous model in the thermodynamic limit, in the absence of external confinement. For the nondipolar (contact interaction only) limit, if intercomponent interactions dominate over the intracomponent interactions, i.e., $g_{12}>\sqrt{g_{11}g_{22}}$, the BECs phase separate into an immiscible ground state \cite{Hall1998B,Papp2008B,McCarron2011,Lercher2011a,Wacker2015,Wang2016,Ho1996,Esry1997,Pu1998,Timmermans1998} (we assume $g_{ii}\ge0$ as the system is unstable otherwise).
In the other extreme, for attractive intercomponent interactions that satisfy $g_{12}<-\sqrt{g_{11}g_{22}}$, the mixture is unstable to implosion, even if both BECs would be individually stable.
This then leaves the intermediate scenario, $g_{12}^2\le g_{11}g_{22}$, as the miscible regime where the two BECs are cospatial and stable.

For dipolar BECs, on the other hand, one can write an analogous homogeneous inequality for the uniform system \cite{Wilson2012a}. For illustration, we take the two miscible components to have equal density ($n_1=n_2$, where $n_i=|\psi_i|^2$), so that the dispersion branches representing the in-phase (density) and out-of-phase (spin) fluctuations take the form
\begin{equation}
    E_\pm^2(\bk)=\frac{\epsilon_1^2+\epsilon_2^2}2 \pm\sqrt{\left(\frac{\epsilon_1^2-\epsilon_2^2}2\right)^2+\left(\frac{n_i\Ut_{12}\hbar^2k^2}{M}\right)^2},\label{homogEpm}
\end{equation}
where
\begin{align}
    \epsilon_i^2=\frac{\hbar^2k^2}{2M}\left(\frac{\hbar^2k^2}{2M}+2\Ut_{ii}n_i\right)
\end{align}
is the single-component uniform system dispersion, and
\begin{equation}
\Ut_{ij}(\theta_k)=g_{ij}+\gdd_{ij}(3\cos^2\theta_k-1)
\end{equation}
is the interaction strength between component $i$ and $j$ in $k$-space along direction $\theta_k$, where $\theta_k$ is the angle between the dipole polarization axis and the wave vector. 

Stability for the homogeneous case can be tested in the long-wavelength limit, $k\to0$, where stability occurs when the spectrum in Eq.~(\ref{homogEpm}) is real.
It should first be noted that stability can only occur if $\edd_{ii}\equiv \gdd_{ii}/g_{ii}\le1$ for both $i=1,2$, to prevent either component from individually imploding.
Stability additionally requires 
\begin{equation}
\Ut_{12}^2 \le \Ut_{11}\Ut_{22}, \label{Eq:homogeneous_inequality}
\end{equation}
a condition that is most stringent when $\theta_k=\pi/2$, yielding the inequality for the coupling constants
\begin{align}
\delta_{12}^2(1-\edd_{12})^2 \le (1-\edd_{11})(1-\edd_{22}) , \label{Eq:ho_ineq_edd}
\end{align}
where 
\begin{equation}
\edd_{ij}=\gdd_{ij}/g_{ij}\label{Eq:rel_edd}
\end{equation}
is the relative dipole strength
and 
\begin{equation}
\delta_{12}=\frac{a_{12}}{\sqrt{a_{11}a_{22}}}=\frac{g_{12}}{\sqrt{g_{11}g_{22}}}\label{Eq:rel_a12}
\end{equation}
is the relative interspecies scattering length, with  the second equality holding as both components have equal masses.
Note that in the absence of dipolar interactions, Eq.~(\ref{Eq:ho_ineq_edd}) immediately reduces to the nondipolar miscibility condition $\delta_{12}^2\le1$, with immiscibility occurring for $1< \delta_{12}$ and mechanical instability when $\delta_{12}<-1$.

Figure~\ref{Fig:figure1} depicts a phase diagram describing Eq.~\eqref{Eq:ho_ineq_edd} for three cases: Fig.~\ref{Fig:figure1}(a), components with parallel dipoles $(\mum_{1}=\mum_{2})$; Fig.~\ref{Fig:figure1}(b), a dipolar and a nondipolar component $(\mum_{1}=0)$; and Fig.~\ref{Fig:figure1}(c), components with antiparallel equal-strength dipoles $(\mum_{1}=-\mum_{2})$.
The axes are the relative interspecies scattering length versus the relative dipole strength.
On the left, all three subplots recover the nondipolar limit where the miscible phase exists for $-1\le\delta_{12}\le1$, with immiscibility occurring above and mechanical instability occurring below.
For all cases, the miscible region shrinks as the system becomes more dipolar, vanishing altogether at the vertical dashed lines, to the right of which even single component dipolar systems are mechanically unstable.

It is intriguing to note that for parallel dipoles [Fig.~\ref{Fig:figure1}(a)], increasing the dipole strength has no affect on the immiscibility threshold, yet the region of mechanical instability steadily grows, eventually pinching off the miscible region. For antiparallel dipoles [Fig.~\ref{Fig:figure1}(c)], the opposite occurs in the sense that dipole strength has no effect on the mechanical instability threshold, but the size of the immiscible region grows. More precisely, Eq.~\eqref{Eq:ho_ineq_edd} gives that for $\mum_{1}=\mum_{2}$ the miscible region is $2\edd_{22}-1\le\delta_{12}\le1$ [Fig.~\ref{Fig:figure1}(a)]; for $\mum_{1}=0$ it is $|\delta_{12}|\le\sqrt{1-\edd_{22}}$ [Fig.~\ref{Fig:figure1}(b)]; while for $\mum_{1}=-\mum_{2}$ the region is $-1\le\delta_{12}\le1-2\edd_{22}$  [Fig.~\ref{Fig:figure1}(c)].

We can categorize the unstable regions as immiscible or mechanically unstable by the character of the unstable excitations in a similar manner to the contact case. To illustrate this in Figs.~\ref{Fig:figure1}(a1--a3), we show the excitation dispersion relation for three points in the phase diagram of  Fig.~\ref{Fig:figure1}(a). In the immiscible regime [Fig.~\ref{Fig:figure1}(a1)], spin (out-of-phase) excitations are dynamically unstable and cause the component densities to spatially separate. Conversely, in the mechanically unstable regime [Fig.~\ref{Fig:figure1}(a3)], density (in-phase) excitations are dynamically unstable.

\begin{figure*}
	\begin{center}
		\centering
		\includegraphics[width=7in]{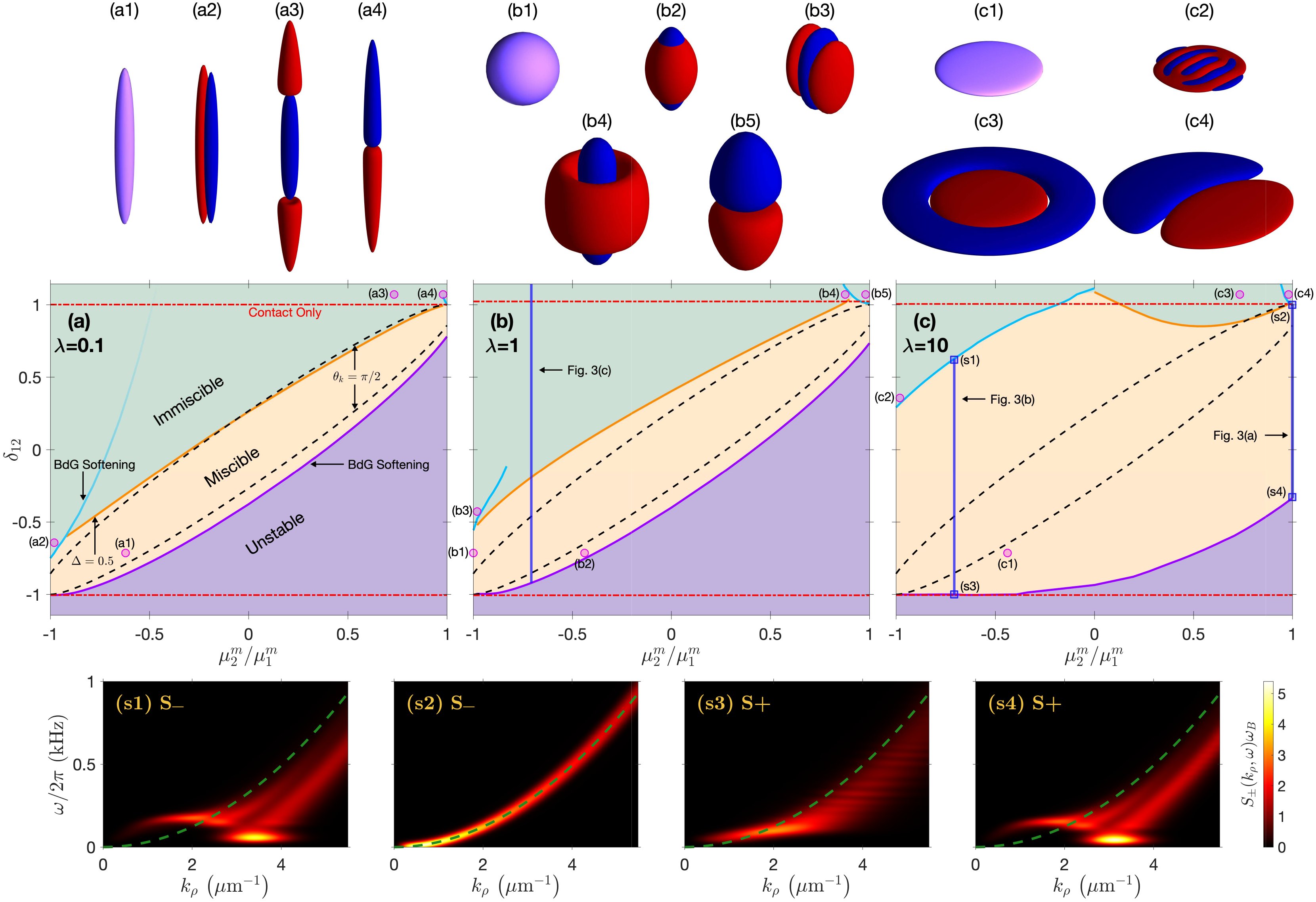}
		\caption{Phase diagrams for cylindrically symmetric (a) cigar-shaped $\lambda=0.1$, (b) spherical $\lambda=1$, (c) and pancake-shaped $\lambda=10$ traps, all with a fixed geometric mean $(\omega_\rho^2 \omega_z)^{1/3}/2\pi=100$\,Hz. The $y$ axis is the relative interspecies contact interaction strength [Eq.~(\ref{Eq:rel_a12})] and the $x$ axis is the relative magnetic dipole moment, with $\mum_2$ varying and $\mum_1= 9.9\mu_B$ fixed, corresponding to $a^{dd}_{11}=130a_0$.
Other parameters are $a_{11}=a_{22}=140a_0$, $N_1=N_2=10^4$.
For all geometries the miscible phase is sandwiched between the immiscible phase above, and a mechanical instability below. The mechanical instability boundary (solid purple lines) occurs when a BdG excitation energy softens to zero, indicating the onset of a dynamic instability, while the immiscibility boundary is characterized either by a dynamic instability (solid blue lines) or a crossover (solid orange lines) to a symmetric immiscible phase. The latter is characterized by the immiscibility contrast $\Delta=0.5$ [Eq.~(\ref{Eq:Contrast})]. For comparison, the homogeneous predictions (\ref{Eq:ho_ineq_edd}) are drawn as dashed lines and the nondipolar calculations are dot-dashed lines.
Isodensity surfaces (top) are plotted at 50\% of the peak density for a given component at the points indicated in the main plots. Component 1 (2) is blue (red), while purple surfaces indicate components with almost overlapping density distributions.
States for a given aspect ratio are drawn to the same scale.
(s1,s2) Spin $S_-(k_\rho,\omega)$ and (s3,s4) density $S_+(k_\rho,\omega)$ dynamic structure factors, with $k_z=0$, for the points marked in (c), showing excitations that have nearly softened to zero at the corresponding immiscibility and instability thresholds. The green dashed lines mark the free particle dispersion. We broadened the $\delta$ function in the dynamic structure factors by setting $\delta(\omega)\approx e^{-(\omega/\omega_B)^2}/\sqrt{\pi}\omega_B$, with $\omega_B=0.5\omega_z$.
}\label{Fig:figure2}
	\end{center}
\end{figure*}

\section{Harmonically trapped mixtures}\label{Sec:Trapped}

Realistic trapping geometries richly interplay with the long-ranged and anisotropic DDIs.
In this section we consider two-component dipolar BECs under 3D confinement, focusing on the role that the two dipole moments---which we allow to be different---have on miscibility and stability.
It should be stressed that the dipoles within a given component are always identical in both strength and orientation.
Our harmonic traps are cylindrically symmetric with the dipoles aligned (or antialigned) along the symmetry axis.
To highlight the role of dipolar interactions, we take both components to have the mass of Dy ($M=164\,\mathrm{u}$), equal populations ($N_1=N_2=10^4$), and balanced intracomponent interactions ($a_{11}=a_{22}=140a_0$).
In Fig.~\ref{Fig:figure2} we present phase diagrams of the immiscible, miscible and unstable regions.
Figures.~\ref{Fig:figure2}(a--c) respectively show cigar-shaped ($\lambda=0.1$), spherical ($\lambda=1$), and pancake-shaped ($\lambda=10$) trapping geometries.
As will be discussed shortly, dipolar interactions generally act to destabilize the miscible phase in the cigar-shaped and spherical geometries. This mirrors what we saw for homogenous mixtures, with parallel dipoles tending to reduce the size of the miscible region in favor of mechanical instability, while antiparallel dipoles instead expand the immiscible region. In contrast, pancake-shaped traps may significantly stabilize the miscible phase in the presence of dipolar interactions.

\subsection{Cigar-shaped trap}

Isosurface density plots are shown for the cigar-shaped trap in
Figs.~\ref{Fig:figure2}(a1--a4), respectively showing examples of miscible, radially immiscible, symmetric immiscible, and asymmetric immiscible ground states.
It is interesting to note how magnetostriction is weakest---with the BEC shape most resembling the trap---for the miscible case when the two dipole moments point in opposite directions [Fig.~\ref{Fig:figure2}(a1)]. This is due to a partial cancellation of the dipolar interaction energy.

The miscible phase is the least stable for the cigar-shaped geometry [Fig.~\ref{Fig:figure2}(a)]; i.e.,~the miscible regime occupies only a narrow region of the phase diagram.
By increasing the intercomponent contact interactions $\delta_{12}$ we soon enter the immiscible regime (green), whereas lowering $\delta_{12}$ causes the miscible phase to become mechanically unstable to implosion (purple).

The mechanical instability boundary (solid purple line) is associated with an $m=0$ excitation energy softening to zero, where $m$ is the angular momentum quantum number.
The immiscibility boundary is qualitatively different, as no excitation energy softens except for small regions where the dipoles are nearly equal strength, i.e.,~$\mum_2/\mum_1 \approx \pm 1$ (solid blue lines).
The immiscibility boundary is instead predominantly a crossover (solid orange line), for which immiscibility occurs without breaking any symmetry [see Fig.~\ref{Fig:figure2}(a3): in this subplot the less dipolar component (red) is pushed out from the trap center].
We quantify immiscibility in these crossover regimes using the contrast equation \cite{Lee2016},
\begin{equation}
\Delta=\left| \frac{n_1(\mathbf{0})}{\max\left\{ n_1(\mathbf{x}) \right\} } - \frac{n_2(\mathbf{0})}{\max\left\{ n_2(\mathbf{x}) \right\} } \right| , \label{Eq:Contrast}
\end{equation}
which provides a strong signal when one component displaces the other from the trap center. The solid orange line marks an immiscibility contrast of $\Delta=0.5$.

For a small region of the immiscibility boundary where the dipoles are antiparallel and nearly equal strength, $\mum_2/\mum_1 \approx -1$, there is a symmetry breaking associated with the softening of an $m=1$ excitation. This suggests an immiscible state with the components arranged side by side.
An even smaller symmetry-breaking region occurs for $\mum_2/\mum_1 \approx 1$, where an $m=0$ mode softening signals an immiscible ground state with the components stacked vertically [Fig.~\ref{Fig:figure2}(a4)].

It is intriguing that for parallel equal-strength dipoles, $\mum_2/\mum_1 = 1$, the immiscibility boundary quantitatively agrees with the nondipolar prediction (horizontal dot-dashed line at $\delta_{12}\approx 1$), but the agreement steadily worsens as the dipoles tend to the antiparallel equal-strength limit $\mum_2/\mum_1 = -1$. In contrast, the instability boundary agrees with the nondipolar result ($\delta_{12}\approx -1$) when the dipoles are antiparallel, and it is worse when they are parallel.
This can be understood from the homogeneous result [Eq.~(\ref{Eq:ho_ineq_edd})], noting in particular that for antiparallel equal-strength dipoles $\gdd_{11} = \gdd_{22} = -\gdd_{12}$.
In fact, for the cigar-shaped geometry the entire phase diagram agrees well with the homogeneous prediction (dashed lines in Fig.~\ref{Fig:figure2}).
This is because both immiscibility and instability are driven by the attractive head-to-tail DDIs, a situation that is unhindered for both the homogeneous limit and cigar-shaped geometries.
The qualitative reduction in size of the miscible region [see Fig.~\ref{Fig:figure2}(a)] compared to the nondipolar prediction ($-1\le  \delta_{12} \le 1$) can then be understood in the following way. When the dipoles are parallel, their mutual head-to-tail DDIs work together to mechanically destabilize the BEC. On the other hand, antiparallel dipoles tend to cancel their DDI energy in the miscible phase, but the energy can be substantially lowered if the components become immiscible.

\subsection{Spherical trap}

Isosurface density contours are displayed for a spherical trap in Figs.~\ref{Fig:figure2}(b1--b5). Interestingly, when the components are miscible and their dipoles are antiparallel and equal strength [Fig.~\ref{Fig:figure2}(b1)], magnetostriction effects cancel and the density distribution is spherical.
Magnetostriction comes back into play once the antiparallel dipoles have unequal magnitudes, elongating the miscible BECs [Fig.~\ref{Fig:figure2}(b2)].

The phase diagram [Fig.~\ref{Fig:figure2}(b)] is qualitatively similar to that for the cigar-shaped trap; however, some differences are apparent. The first is that deviations from the homogeneous result (dashed lines) are larger.
The increased stability of the miscible regime, i.e., the increased area on the phase diagram, arises from the tighter trap along the direction of dipole orientation, which partially disrupts the destabilizing head-to-tail dipolar attraction.
Another difference is that a cusp is now prominent along the immiscibility boundary near the parallel equal-strength dipole limit $\mum_2/\mum_1 \approx 1$ [top right of Fig.~\ref{Fig:figure2}(b)].
This is associated with a symmetry-breaking phase transition between symmetric-immiscible [Fig.~\ref{Fig:figure2}(b4)] and asymmetric-immiscible phases [Fig.~\ref{Fig:figure2}(b5)].
Note that for the parameters in Fig.~\ref{Fig:figure2}(b4), it is the less-dipolar component (red) that is wrapped around the inner component.
In Sec.~\ref{Sec:SAIP}, we will return in detail to the concept of the symmetric-immiscible to asymmetric-immiscible phase transition.

\subsection{Pancake-shaped trap}

The phase diagram for the pancake-shaped trap [Fig.~\ref{Fig:figure2}(c)] is qualitatively different from the other trapping geometries, and the homogeneous predictions are less useful.
The miscible region is substantially enlarged, both to lower and higher $\delta_{12}$, thanks to the tight confinement along the direction of dipole polarization, $z$, which significantly disrupts the head-to-tail attraction between dipoles.
However, it is important to note that the immiscible boundary still agrees with the nondipolar case when $\mum_2/\mum_1=1$, as well as there being analogous agreement for the instability boundary when $\mum_2/\mum_1=-1$, due to the cancellation of dipolar effects for spin and density excitations, respectively.

We first consider the immiscibility boundary, and to gain insight, we plot the dynamic structure factor for two points [see Figs.~\ref{Fig:figure2}(s1--s2)].
These subplots display the spin structure factor [$S_-$ in Eq.~\eqref{Eq.SF}], as the spin excitations are the ones responsible for relative motion between the components and hence also immiscibility.
In Fig.~\ref{Fig:figure2}(s1), the softening of a roton mode is clearly visible, suggesting an immiscible ground state with alternating domains within the $x$-$y$ plane.
In fact, a roton immiscibility phase transition (solid blue line) spans approximately half the phase diagram, i.e.,~$-1\leq\mum_2/\mum_1\lesssim 0$.
For reference, the roton immiscibility for the quasi-two-dimensional (2D) system discussed in \cite{Wilson2012a} was for the case $\mum_2/\mum_1=0$.
In contrast, the region of the phase diagram spanning $0\lesssim\mum_2/\mum_1\lesssim1$ is again dominated by a crossover to a symmetric-immiscible phase (solid orange line).
Interestingly, Fig.~\ref{Fig:figure2}(s2) shows a noninteracting dispersion relation for $\mum_2/\mum_1=1$.
This not only demonstrates a cancellation of the DDI effects for the immiscibility boundary, but also implies a nullification of the quantum spin fluctuations, which is also expected for nondipolar mixtures when $\delta_{12}=1$ \cite{Bisset2018}.

As we saw for the other geometries, the mechanical instability boundary for the pancake-shaped trap is once again characterized by BdG excitations softening to zero across the entire width of the phase diagram.
Since instability occurs with both components moving in phase, Figs.~\ref{Fig:figure2}(s3--s4) display the density structure factor, i.e.,~$S_+$ in Eq.~\eqref{Eq.SF}.
Figure~\ref{Fig:figure2}(s3) shows that near the antiparallel equal-strength limit $\mum_2/\mum_1\approx-1$, the dispersion relation shows less roton softening, and will again approach a noninteracting form. Moving towards the parallel dipole configuration, Fig.~\ref{Fig:figure2}(s4) demonstrates that the mechanical instability phase transition is again driven by rotons. Here, both components prefer to move in phase, and this roton is analogous to the original one envisioned for single-component dipolar BECs \cite{Santos2003a,ODell2003a,Chomaz2018a,Petter2019}.

\begin{figure*}
	\begin{center}
        \includegraphics[width=5.5in]{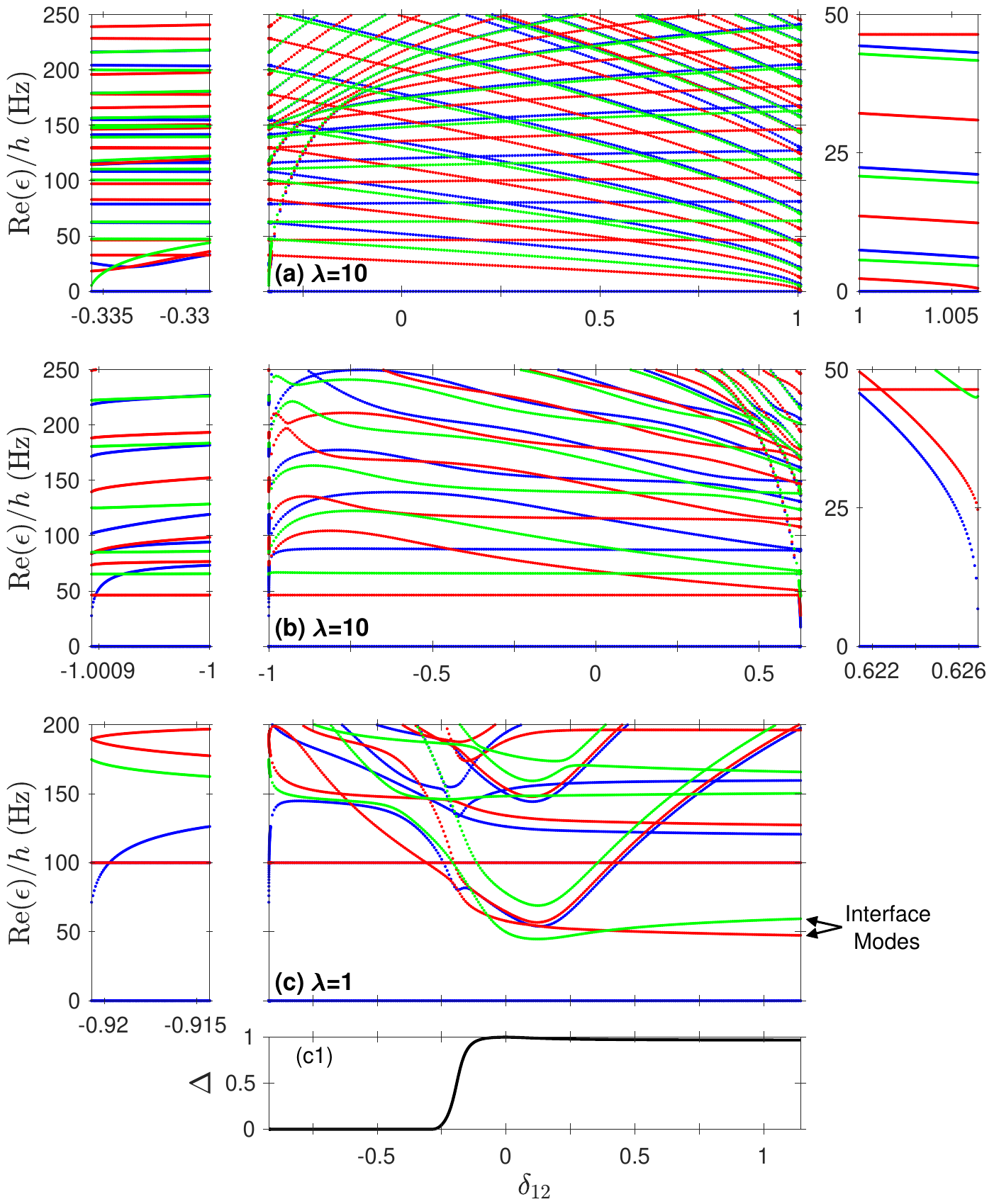}
		\caption{BdG excitation energies for the three vertical trajectories drawn in Fig.~\ref{Fig:figure2}, as a function of the relative interspecies contact interaction
			strength (\ref{Eq:rel_a12}).
(a, b) pancake-shaped trap with $\mum_1/\mum_2=\{1,-1/\sqrt2\}$, respectively, and (c) spherical trap with $\mum_1/\mum_2=-1/\sqrt2$.
The blue, red, and green lines show excitations with angular momentum quantum numbers $m=0,1$, and $2$, respectively (we neglect $m>2$ for clarity).
The subplots to the sides are zoomed-in portions of the central figures to highlight excitation softenings.
In (c1) we plot the immiscibility contrast $\Delta$ to highlight the miscible to symmetric-immiscible crossover for the spherical case, which sees no excitations soften to zero.}\label{Fig:figure3}
	\end{center}
\end{figure*}

\subsection{The role of excitations for instability and immiscibility}

To gain further insight into the various phase transitions and crossovers, Fig.~\ref{Fig:figure3} displays the BdG excitation energies along the three vertical trajectories drawn in Figs.~\ref{Fig:figure2}(b) and~\ref{Fig:figure2}(c). Figures~\ref{Fig:figure3}(a) and~\ref{Fig:figure3}(b) represent two cuts through the miscible regime for the pancake-shaped trap. For both, the stability of the miscible phase is disrupted by a BdG energy softening on either side: density modes for the mechanical instability on the left, and spin modes for the transition to immiscibility on the right.
While approaching the instability to the left of Fig.~\ref{Fig:figure3}(b), it is interesting that the energies generally appear to increase, giving little warning for the sudden softening and the impending collapse.

The case of Fig.~\ref{Fig:figure3}(c) is for the spherical trap considered in Fig.~\ref{Fig:figure2}(b).
While an excitation energy also suddenly dives on approach to the instability on the left, no such softening is exhibited for the immiscibility on the right, and no symmetry is broken.
Instead, immiscibility occurs as a crossover characterized in Fig.~\ref{Fig:figure3}(c1) by a rapidly rising immiscibility contrast $\Delta$ [Eq.~(\ref{Eq:Contrast})].
Although no BdG energies touch zero, many do exhibit a dip in energy at $\delta_{12}\approx 0.15$.
In the immiscible phase, the role of the spin excitations is reversed, instead acting to mix the separated components, which becomes increasingly costly from an energy perspective as $\delta_{12}$ rises beyond the dip.
The excitations that remain at low energy for increasing $\delta_{12}$ represent excitations of the interface separating the components.

\begin{figure}
	\begin{center}
        \includegraphics[width=3.45in]{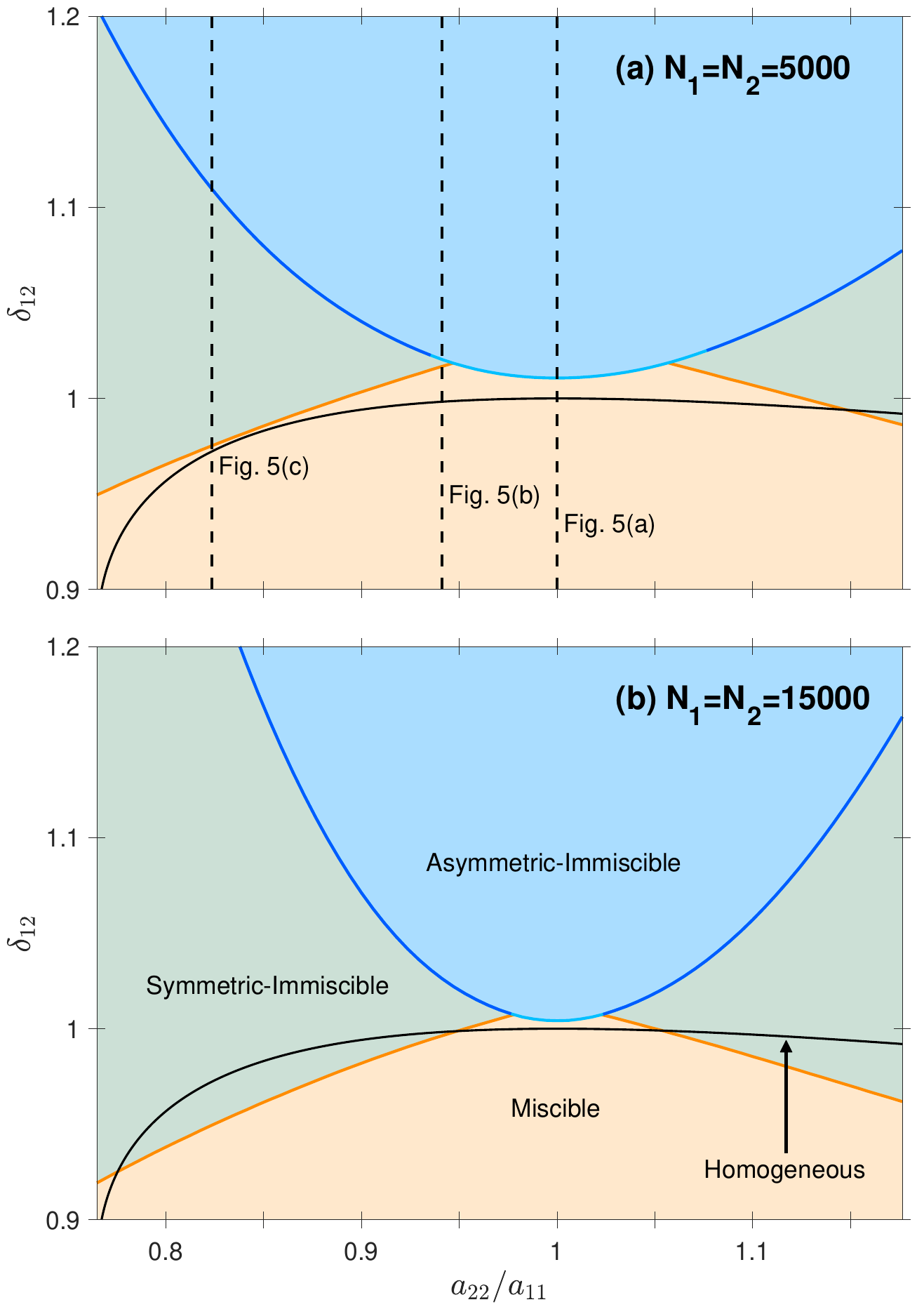}
		\caption{Ground-state phase diagram of the miscible (orange), symmetric-immiscible (green), and asymmetric-immiscible (blue) states.
The solid blue lines mark the symmetry-breaking transition to the asymmetric-immiscible phase, which can be continuous (the states in the two phases are the same at the boundary, light blue) or first order (dark blue). The solid orange lines indicate the immiscibility contrast $\Delta=0.5$ [Eq.~(\ref{Eq:Contrast})].
(a) $N_1=N_2=5000$ and (b) $N_1=N_2=15000$.
Other parameters: The dipoles are parallel with $a^{dd}_{11}=a^{dd}_{22}=130\,a_0,~M=164\,\mathrm{u},~\omega_\rho/2\pi=60\,\mathrm{Hz},$ and $~\omega_z/2\pi=30\,\mathrm{Hz}$. We keep $a_{11}=170a_0$ fixed and study the effects of varying $a_{22}$ and the relative interspecies scattering length, $\delta_{12}$.
The vertical dashed lines indicate trajectories considered in Fig.~\ref{Fig:figure5}.
}\label{Fig:figure4}
	\end{center}
\end{figure}

\begin{figure*}
	\begin{center}
		\includegraphics[width=7in]{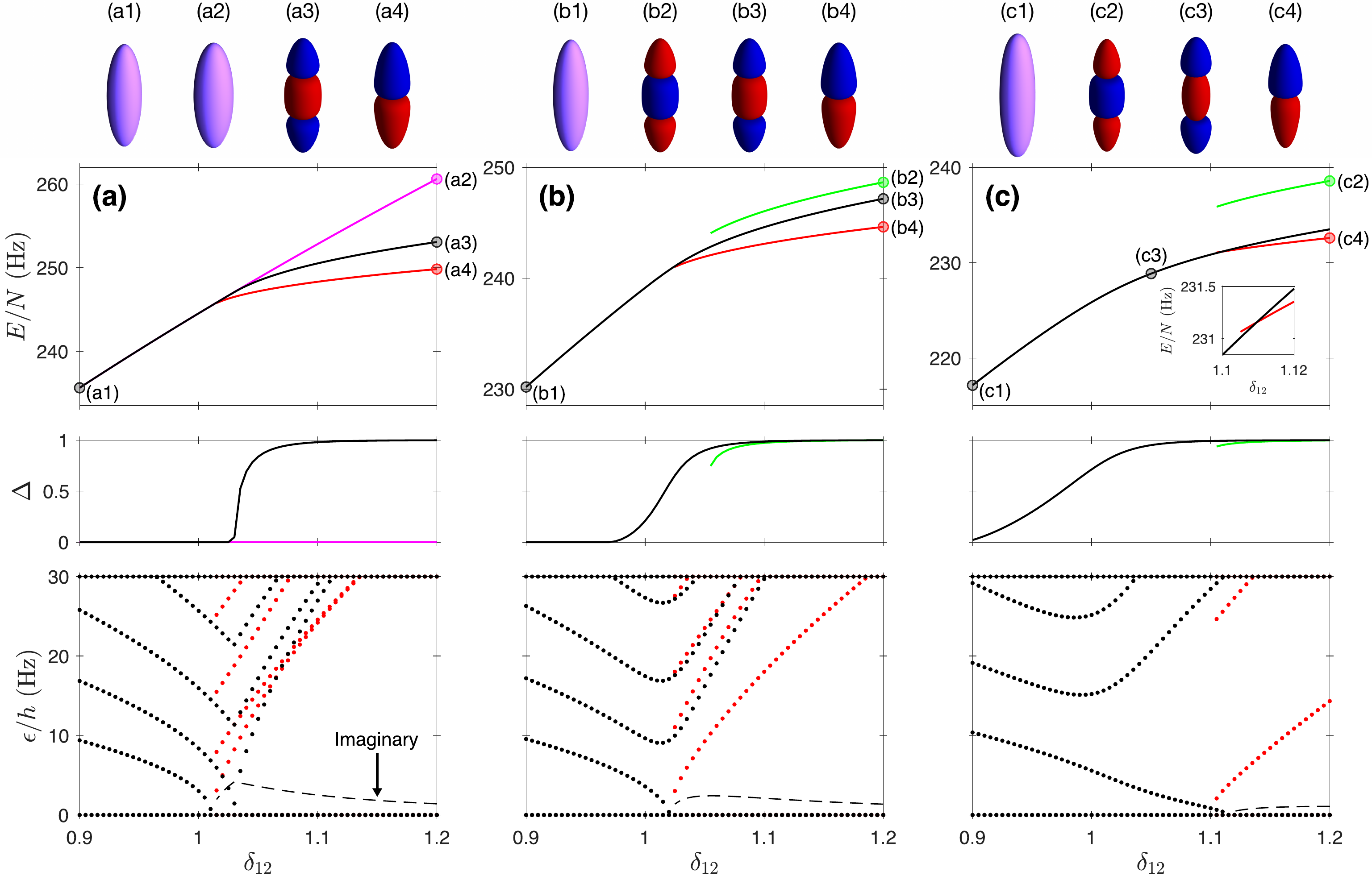}
		\caption{Stationary state properties along the three trajectories drawn as dashed vertical lines in Fig.~\ref{Fig:figure4}(a). From top to bottom the rows are isodensity surface contours, stationary-state energies, immiscibility contrast $\Delta$, and BdG excitation energies of the lowest two branches (the lowest three rows of subplots are color coordinated: $\Delta$ is inappropriate for the asymmetric case and not shown). These quantities are plotted versus the relative interspecies scattering length $\delta_{12}$. The columns are (a)  $a_{22}=170a_0$, (b) $a_{22}=160a_0$, and (c) $a_{22}=140a_0$. The isodensity surfaces are drawn at 50\% of the respective peak densities for components 1 (blue) and 2 (red). For clarity, the lower panels only show $m=0$ excitations as these are most relevant for this regime. Other parameters are $N_1=N_2=5000$, $\omega_\rho/2\pi=60\mathrm{Hz},~\omega_z/2\pi=30\mathrm{Hz}$, and~$a_{11}=170a_0$, and the dipoles are parallel with $a^{dd}_{11}=a^{dd}_{22}=130a_0$.
}\label{Fig:figure5}
	\end{center}
\end{figure*}

\section{Symmetric-immiscible and asymmetric-immiscible phases} \label{Sec:SAIP}

\subsection{Phase diagram}

In the previous section, we saw a symmetry-breaking phase transition between symmetric-immiscible and asymmetric-immiscible states [see, e.g.,~Figs.~\ref{Fig:figure2}(a3--a4) or Figs.~\ref{Fig:figure2}(b4--b5)].
In this section we further investigate such states, finding that although both are generally important, the direct transition between miscible and asymmetric-immiscible ground states only occurs for nearly balanced mixtures.

A phase diagram is presented in Fig.~\ref{Fig:figure4} for a mildly oblate trap ($\lambda=0.5$) with balanced atom numbers and dipoles, $\mum_2=\mum_1$, displaying miscible (lower), asymmetric-immiscible (upper middle) and symmetric-immiscible ground-state regions. The balance between the components is only broken by the intraspecies scattering lengths $a_{22}/a_{11}$ displayed along the $x$ axis.

Figure~\ref{Fig:figure4}(a) considers relatively small populations $N_1=N_2=5000$.
As expected, the miscible phase occurs at low $\delta_{12}$. For a sizable region surrounding the balanced situation, $a_{22}/a_{11}=1$, the ground state directly transitions to an asymmetric-immiscible phase. However, larger atom numbers ($N_1=N_2=15000$) are considered in Fig.~\ref{Fig:figure4}(b), and here we can already see a substantial reduction in the length of this symmetry-breaking miscible-immiscible phase transition.
This can be understood by considering the kinetic energy cost of domain walls.
For small $N$, the asymmetric-immiscible phase is more likely favored as this lowers the domain wall area. But for large $N$, the interface energy becomes less important, and any asymmetry between the components---such as population or intraspecies interactions---will favor one component being positioned at the center, resulting in a symmetric-immiscible state.
As $\delta_{12}$ increases, however, interface energy can again become dominant and a symmetric-immiscible ground state may eventually break symmetry, as evidenced by the increasing width seen in Figs.~\ref{Fig:figure4}(a) and~\ref{Fig:figure4}(b).

\subsection{Symmetric and asymmetric branches}

In Fig.~\ref{Fig:figure5} we explore in detail the various symmetric and asymmetric stationary states along the three trajectories marked as vertical dashed lines in Fig.~\ref{Fig:figure4}(a).
On the left, Fig.~\ref{Fig:figure5}(a) demonstrates that when the interactions and populations are balanced, the change from miscibility [Fig.~\ref{Fig:figure5}(a1)] to immiscibility occurs directly via a second-order phase transition to an asymmetric-immiscible phase [Fig.~\ref{Fig:figure5}(a4)]. The miscible branch continues to higher intercomponent interactions $\delta_{12}$, but is dynamically unstable [Fig.~\ref{Fig:figure5}(a2)]. Likewise, the symmetric-immiscible branch [Fig.~\ref{Fig:figure5}(a3)]---which is doubly degenerate here since the components can be interchanged at no energy cost---is also unstable, indicated by the imaginary excitation energy in the lower subplot.

Once the components become imbalanced, $\{a_{11},a_{22}\}=\{170,160\}a_0$, Fig.~\ref{Fig:figure5}(b) shows that the degeneracy of the symmetric-immiscible modes [Figs.~\ref{Fig:figure5}(b2--b3)] is broken. This arises from the energetic advantage of having the component with weaker interactions (red) occupy the trap center [Fig.~\ref{Fig:figure5}(b3)]. Interestingly, the upper branch [Fig.~\ref{Fig:figure5}(b2)] disconnects from the others. However, despite these interesting developments, the ground state still undergoes a phase transition directly from the miscible to asymmetric-immiscible phase.

Figure~\ref{Fig:figure5}(c) demonstrates that once the imbalance is large enough---here, \{$a_{11},a_{22}$\}$\,=\,$\{$170,140$\}$a_0$---a region of symmetric-immiscible phase develops between the miscible and asymmetric-immiscible phases.
The miscible-immiscible boundary now becomes crossover, as can be seen by observing Fig.~\ref{Fig:figure5}(c3), as well as noticing how the immiscibility contrast $\Delta$ grows long before the other stationary-state branches appear.
The symmetry does eventually break just above $\delta_{12}=1.1$, however, resulting in a first-order phase transition between the symmetric-immiscible and asymmetric-immiscible phases. The first-order character is apparent as the two stationary-state energies now cross, instead of connecting, and the BdG spectrum confirms a small region of mutual metastability and hence a discontinuity.

\section{Conclusions}\label{Sec:Conc}

We performed an in-depth investigation into binary dipolar BECs, finding rich phenomena that are inaccessible for nondipolar mixtures.
By constructing phase diagrams for various trapping geometries---with the dipoles aligned with the trap's symmetry axis---we demonstrated that the miscible regime is substantially reduced for cigar-shaped and spherical traps, but remains large for pancake-shaped traps.
While the analytic homogeneous prediction does a good job at estimating the immiscibility and mechanical instability thresholds for cigar traps, and a reasonable job for spherical traps, it qualitatively fails for the pancake geometries.

Our phase diagrams considered two components with a wide range of dipole combinations, ranging in a continuous way from the parallel to the antiparallel situation.
Intriguingly, components with parallel dipoles tend to cancel one another's influence on immiscibility, but strongly work together to trigger mechanical instability. The opposite occurs for antiparallel dipoles, which tend to cancel their effect on instability while maximally affecting immiscibility.

By performing BdG calculations we demonstrated that for certain regions of the phase diagrams, immiscibility is triggered by an excitation energy softening to zero, whereas for other regions there is instead a crossover with no symmetry breaking. In contrast, the mechanical instability boundary is always driven by a BdG softening for all regimes that we considered.
For the pancake-shaped geometry we found a broad region of the phase diagram dominated by a roton instability and another dominated by a roton immiscibility.
While this work focused on the regime where beyond meanfield quantum fluctuations are qualitatively unimportant, an interesting future direction will be to extend our analyses into the droplet regime for different trapping geometries and dipole combinations (see related work for untrapped self-bound droplets \cite{Bisset2021,Smith2021}).

\noindent {\bf Acknowledgments: $\,$}
We would like to thank Tom Bienaim\'e, Lauriane Chomaz, Francesca Ferlaino, Sukla Pal, and Joseph Smith for stimulating discussions.
D.B. and P.B.B. were supported by the Marsden Fund of New Zealand.



\begin{thebibliography}{53}%
\makeatletter
\providecommand \@ifxundefined [1]{%
 \@ifx{#1\undefined}
}%
\providecommand \@ifnum [1]{%
 \ifnum #1\expandafter \@firstoftwo
 \else \expandafter \@secondoftwo
 \fi
}%
\providecommand \@ifx [1]{%
 \ifx #1\expandafter \@firstoftwo
 \else \expandafter \@secondoftwo
 \fi
}%
\providecommand \natexlab [1]{#1}%
\providecommand \enquote  [1]{``#1''}%
\providecommand \bibnamefont  [1]{#1}%
\providecommand \bibfnamefont [1]{#1}%
\providecommand \citenamefont [1]{#1}%
\providecommand \href@noop [0]{\@secondoftwo}%
\providecommand \href [0]{\begingroup \@sanitize@url \@href}%
\providecommand \@href[1]{\@@startlink{#1}\@@href}%
\providecommand \@@href[1]{\endgroup#1\@@endlink}%
\providecommand \@sanitize@url [0]{\catcode `\\12\catcode `\$12\catcode
  `\&12\catcode `\#12\catcode `\^12\catcode `\_12\catcode `\%12\relax}%
\providecommand \@@startlink[1]{}%
\providecommand \@@endlink[0]{}%
\providecommand \url  [0]{\begingroup\@sanitize@url \@url }%
\providecommand \@url [1]{\endgroup\@href {#1}{\urlprefix }}%
\providecommand \urlprefix  [0]{URL }%
\providecommand \Eprint [0]{\href }%
\providecommand \doibase [0]{http://dx.doi.org/}%
\providecommand \selectlanguage [0]{\@gobble}%
\providecommand \bibinfo  [0]{\@secondoftwo}%
\providecommand \bibfield  [0]{\@secondoftwo}%
\providecommand \translation [1]{[#1]}%
\providecommand \BibitemOpen [0]{}%
\providecommand \bibitemStop [0]{}%
\providecommand \bibitemNoStop [0]{.\EOS\space}%
\providecommand \EOS [0]{\spacefactor3000\relax}%
\providecommand \BibitemShut  [1]{\csname bibitem#1\endcsname}%
\let\auto@bib@innerbib\@empty
\bibitem [{\citenamefont {Hall}\ \emph {et~al.}(1998)\citenamefont {Hall},
  \citenamefont {Matthews}, \citenamefont {Ensher}, \citenamefont {Wieman},\
  and\ \citenamefont {Cornell}}]{Hall1998B}%
  \BibitemOpen
  \bibfield  {author} {\bibinfo {author} {\bibfnamefont {D.~S.}\ \bibnamefont
  {Hall}}, \bibinfo {author} {\bibfnamefont {M.~R.}\ \bibnamefont {Matthews}},
  \bibinfo {author} {\bibfnamefont {J.~R.}\ \bibnamefont {Ensher}}, \bibinfo
  {author} {\bibfnamefont {C.~E.}\ \bibnamefont {Wieman}}, \ and\ \bibinfo
  {author} {\bibfnamefont {E.~A.}\ \bibnamefont {Cornell}},\ }\href {\doibase %
  10.1103/PhysRevLett.81.1539} {\bibfield  {journal} {\bibinfo  {journal}
  {Phys. Rev. Lett.}\ }\textbf {\bibinfo {volume} {81}},\ \bibinfo {pages}
  {1539} (\bibinfo {year} {1998})}\BibitemShut {NoStop}%
\bibitem [{\citenamefont {Papp}\ \emph {et~al.}(2008)\citenamefont {Papp},
  \citenamefont {Pino},\ and\ \citenamefont {Wieman}}]{Papp2008B}%
  \BibitemOpen
  \bibfield  {author} {\bibinfo {author} {\bibfnamefont {S.~B.}\ \bibnamefont
  {Papp}}, \bibinfo {author} {\bibfnamefont {J.~M.}\ \bibnamefont {Pino}}, \
  and\ \bibinfo {author} {\bibfnamefont {C.~E.}\ \bibnamefont {Wieman}},\
  }\href {\doibase 10.1103/PhysRevLett.101.040402} {\bibfield  {journal}
  {\bibinfo  {journal} {Phys. Rev. Lett.}\ }\textbf {\bibinfo {volume} {101}},\
  \bibinfo {pages} {040402} (\bibinfo {year} {2008})}\BibitemShut {NoStop}%
\bibitem [{\citenamefont {McCarron}\ \emph {et~al.}(2011)\citenamefont
  {McCarron}, \citenamefont {Cho}, \citenamefont {Jenkin}, \citenamefont
  {K\"oppinger},\ and\ \citenamefont {Cornish}}]{McCarron2011}%
  \BibitemOpen
  \bibfield  {author} {\bibinfo {author} {\bibfnamefont {D.~J.}\ \bibnamefont
  {McCarron}}, \bibinfo {author} {\bibfnamefont {H.~W.}\ \bibnamefont {Cho}},
  \bibinfo {author} {\bibfnamefont {D.~L.}\ \bibnamefont {Jenkin}}, \bibinfo
  {author} {\bibfnamefont {M.~P.}\ \bibnamefont {K\"oppinger}}, \ and\ \bibinfo
  {author} {\bibfnamefont {S.~L.}\ \bibnamefont {Cornish}},\ }\href {\doibase %
  10.1103/PhysRevA.84.011603} {\bibfield  {journal} {\bibinfo  {journal} {Phys.
  Rev. A}\ }\textbf {\bibinfo {volume} {84}},\ \bibinfo {pages} {011603(R)}
  (\bibinfo {year} {2011})}\BibitemShut {NoStop}%
\bibitem [{\citenamefont {Lercher}\ \emph {et~al.}(2011)\citenamefont
  {Lercher}, \citenamefont {Takekoshi}, \citenamefont {Debatin}, \citenamefont
  {Schuster}, \citenamefont {Rameshan}, \citenamefont {Ferlaino}, \citenamefont
  {Grimm},\ and\ \citenamefont {N{\"a}gerl}}]{Lercher2011a}%
  \BibitemOpen
  \bibfield  {author} {\bibinfo {author} {\bibfnamefont {A.}~\bibnamefont
  {Lercher}}, \bibinfo {author} {\bibfnamefont {T.}~\bibnamefont {Takekoshi}},
  \bibinfo {author} {\bibfnamefont {M.}~\bibnamefont {Debatin}}, \bibinfo
  {author} {\bibfnamefont {B.}~\bibnamefont {Schuster}}, \bibinfo {author}
  {\bibfnamefont {R.}~\bibnamefont {Rameshan}}, \bibinfo {author}
  {\bibfnamefont {F.}~\bibnamefont {Ferlaino}}, \bibinfo {author}
  {\bibfnamefont {R.}~\bibnamefont {Grimm}}, \ and\ \bibinfo {author}
  {\bibfnamefont {H.}~\bibnamefont {N{\"a}gerl}},\ }\href {\doibase %
  10.1140/epjd/e2011-20015-6} {\bibfield  {journal} {\bibinfo  {journal} {Eur.
  Phys. J. D}\ }\textbf {\bibinfo {volume} {65}},\ \bibinfo {pages} {3}
  (\bibinfo {year} {2011})}\BibitemShut {NoStop}%
\bibitem [{\citenamefont {Wacker}\ \emph {et~al.}(2015)\citenamefont {Wacker},
  \citenamefont {J\o{}rgensen}, \citenamefont {Birkmose}, \citenamefont
  {Horchani}, \citenamefont {Ertmer}, \citenamefont {Klempt}, \citenamefont
  {Winter}, \citenamefont {Sherson},\ and\ \citenamefont {Arlt}}]{Wacker2015}%
  \BibitemOpen
  \bibfield  {author} {\bibinfo {author} {\bibfnamefont {L.}~\bibnamefont
  {Wacker}}, \bibinfo {author} {\bibfnamefont {N.~B.}\ \bibnamefont
  {J\o{}rgensen}}, \bibinfo {author} {\bibfnamefont {D.}~\bibnamefont
  {Birkmose}}, \bibinfo {author} {\bibfnamefont {R.}~\bibnamefont {Horchani}},
  \bibinfo {author} {\bibfnamefont {W.}~\bibnamefont {Ertmer}}, \bibinfo
  {author} {\bibfnamefont {C.}~\bibnamefont {Klempt}}, \bibinfo {author}
  {\bibfnamefont {N.}~\bibnamefont {Winter}}, \bibinfo {author} {\bibfnamefont
  {J.}~\bibnamefont {Sherson}}, \ and\ \bibinfo {author} {\bibfnamefont
  {J.~J.}\ \bibnamefont {Arlt}},\ }\href {\doibase 10.1103/PhysRevA.92.053602}
  {\bibfield  {journal} {\bibinfo  {journal} {Phys. Rev. A}\ }\textbf {\bibinfo
  {volume} {92}},\ \bibinfo {pages} {053602} (\bibinfo {year}
  {2015})}\BibitemShut {NoStop}%
\bibitem [{\citenamefont {Wang}\ \emph {et~al.}(2016)\citenamefont {Wang},
  \citenamefont {Li}, \citenamefont {Xiong},\ and\ \citenamefont
  {Wang}}]{Wang2016}%
  \BibitemOpen
  \bibfield  {author} {\bibinfo {author} {\bibfnamefont {F.}~\bibnamefont
  {Wang}}, \bibinfo {author} {\bibfnamefont {X.}~\bibnamefont {Li}}, \bibinfo
  {author} {\bibfnamefont {D.}~\bibnamefont {Xiong}}, \ and\ \bibinfo {author}
  {\bibfnamefont {D.}~\bibnamefont {Wang}},\ }\href
  {http://stacks.iop.org/0953-4075/49/i=1/a=015302} {\bibfield  {journal}
  {\bibinfo  {journal} {J. Phys. B}\ }\textbf {\bibinfo {volume} {49}},\
  \bibinfo {pages} {015302} (\bibinfo {year} {2016})}\BibitemShut {NoStop}%
\bibitem [{\citenamefont {Ho}\ and\ \citenamefont {Shenoy}(1996)}]{Ho1996}%
  \BibitemOpen
  \bibfield  {author} {\bibinfo {author} {\bibfnamefont {T.-L.}\ \bibnamefont
  {Ho}}\ and\ \bibinfo {author} {\bibfnamefont {V.~B.}\ \bibnamefont
  {Shenoy}},\ }\href {\doibase 10.1103/PhysRevLett.77.3276} {\bibfield
  {journal} {\bibinfo  {journal} {Phys. Rev. Lett.}\ }\textbf {\bibinfo
  {volume} {77}},\ \bibinfo {pages} {3276} (\bibinfo {year}
  {1996})}\BibitemShut {NoStop}%
\bibitem [{\citenamefont {Esry}\ \emph {et~al.}(1997)\citenamefont {Esry},
  \citenamefont {Greene}, \citenamefont {Burke},\ and\ \citenamefont
  {Bohn}}]{Esry1997}%
  \BibitemOpen
  \bibfield  {author} {\bibinfo {author} {\bibfnamefont {B.~D.}\ \bibnamefont
  {Esry}}, \bibinfo {author} {\bibfnamefont {C.~H.}\ \bibnamefont {Greene}},
  \bibinfo {author} {\bibfnamefont {J.~P.}\ \bibnamefont {Burke}, \bibfnamefont
  {Jr.}}, \ and\ \bibinfo {author} {\bibfnamefont {J.~L.}\ \bibnamefont
  {Bohn}},\ }\href {\doibase 10.1103/PhysRevLett.78.3594} {\bibfield  {journal}
  {\bibinfo  {journal} {Phys. Rev. Lett.}\ }\textbf {\bibinfo {volume} {78}},\
  \bibinfo {pages} {3594} (\bibinfo {year} {1997})}\BibitemShut {NoStop}%
\bibitem [{\citenamefont {Pu}\ and\ \citenamefont {Bigelow}(1998)}]{Pu1998}%
  \BibitemOpen
  \bibfield  {author} {\bibinfo {author} {\bibfnamefont {H.}~\bibnamefont
  {Pu}}\ and\ \bibinfo {author} {\bibfnamefont {N.~P.}\ \bibnamefont
  {Bigelow}},\ }\href {\doibase 10.1103/PhysRevLett.80.1130} {\bibfield
  {journal} {\bibinfo  {journal} {Phys. Rev. Lett.}\ }\textbf {\bibinfo
  {volume} {80}},\ \bibinfo {pages} {1130} (\bibinfo {year}
  {1998})}\BibitemShut {NoStop}%
\bibitem [{\citenamefont {Timmermans}(1998)}]{Timmermans1998}%
  \BibitemOpen
  \bibfield  {author} {\bibinfo {author} {\bibfnamefont {E.}~\bibnamefont
  {Timmermans}},\ }\href {\doibase 10.1103/PhysRevLett.81.5718} {\bibfield
  {journal} {\bibinfo  {journal} {Phys. Rev. Lett.}\ }\textbf {\bibinfo
  {volume} {81}},\ \bibinfo {pages} {5718} (\bibinfo {year}
  {1998})}\BibitemShut {NoStop}%
\bibitem [{\citenamefont {Petrov}(2015)}]{Petrov2015a}%
  \BibitemOpen
  \bibfield  {author} {\bibinfo {author} {\bibfnamefont {D.~S.}\ \bibnamefont
  {Petrov}},\ }\href {\doibase 10.1103/PhysRevLett.115.155302} {\bibfield
  {journal} {\bibinfo  {journal} {Phys. Rev. Lett.}\ }\textbf {\bibinfo
  {volume} {115}},\ \bibinfo {pages} {155302} (\bibinfo {year}
  {2015})}\BibitemShut {NoStop}%
\bibitem [{\citenamefont {Cabrera}\ \emph {et~al.}(2018)\citenamefont
  {Cabrera}, \citenamefont {Tanzi}, \citenamefont {Sanz}, \citenamefont
  {Naylor}, \citenamefont {Thomas}, \citenamefont {Cheiney},\ and\
  \citenamefont {Tarruell}}]{Cabrera2018}%
  \BibitemOpen
  \bibfield  {author} {\bibinfo {author} {\bibfnamefont {C.~R.}\ \bibnamefont
  {Cabrera}}, \bibinfo {author} {\bibfnamefont {L.}~\bibnamefont {Tanzi}},
  \bibinfo {author} {\bibfnamefont {J.}~\bibnamefont {Sanz}}, \bibinfo {author}
  {\bibfnamefont {B.}~\bibnamefont {Naylor}}, \bibinfo {author} {\bibfnamefont
  {P.}~\bibnamefont {Thomas}}, \bibinfo {author} {\bibfnamefont
  {P.}~\bibnamefont {Cheiney}}, \ and\ \bibinfo {author} {\bibfnamefont
  {L.}~\bibnamefont {Tarruell}},\ }\href {\doibase 10.1126/science.aao5686}
  {\bibfield  {journal} {\bibinfo  {journal} {Science}\ }\textbf {\bibinfo
  {volume} {359}},\ \bibinfo {pages} {301} (\bibinfo {year}
  {2018})}\BibitemShut {NoStop}%
\bibitem [{\citenamefont {Semeghini}\ \emph {et~al.}(2018)\citenamefont
  {Semeghini}, \citenamefont {Ferioli}, \citenamefont {Masi}, \citenamefont
  {Mazzinghi}, \citenamefont {Wolswijk}, \citenamefont {Minardi}, \citenamefont
  {Modugno}, \citenamefont {Modugno}, \citenamefont {Inguscio},\ and\
  \citenamefont {Fattori}}]{Semeghini2018}%
  \BibitemOpen
  \bibfield  {author} {\bibinfo {author} {\bibfnamefont {G.}~\bibnamefont
  {Semeghini}}, \bibinfo {author} {\bibfnamefont {G.}~\bibnamefont {Ferioli}},
  \bibinfo {author} {\bibfnamefont {L.}~\bibnamefont {Masi}}, \bibinfo {author}
  {\bibfnamefont {C.}~\bibnamefont {Mazzinghi}}, \bibinfo {author}
  {\bibfnamefont {L.}~\bibnamefont {Wolswijk}}, \bibinfo {author}
  {\bibfnamefont {F.}~\bibnamefont {Minardi}}, \bibinfo {author} {\bibfnamefont
  {M.}~\bibnamefont {Modugno}}, \bibinfo {author} {\bibfnamefont
  {G.}~\bibnamefont {Modugno}}, \bibinfo {author} {\bibfnamefont
  {M.}~\bibnamefont {Inguscio}}, \ and\ \bibinfo {author} {\bibfnamefont
  {M.}~\bibnamefont {Fattori}},\ }\href {\doibase %
  10.1103/PhysRevLett.120.235301} {\bibfield  {journal} {\bibinfo  {journal}
  {Phys. Rev. Lett.}\ }\textbf {\bibinfo {volume} {120}},\ \bibinfo {pages}
  {235301} (\bibinfo {year} {2018})}\BibitemShut {NoStop}%
\bibitem [{\citenamefont {Trautmann}\ \emph {et~al.}(2018)\citenamefont
  {Trautmann}, \citenamefont {Ilzh\"ofer}, \citenamefont {Durastante},
  \citenamefont {Politi}, \citenamefont {Sohmen}, \citenamefont {Mark},\ and\
  \citenamefont {Ferlaino}}]{Trautmann2018}%
  \BibitemOpen
  \bibfield  {author} {\bibinfo {author} {\bibfnamefont {A.}~\bibnamefont
  {Trautmann}}, \bibinfo {author} {\bibfnamefont {P.}~\bibnamefont
  {Ilzh\"ofer}}, \bibinfo {author} {\bibfnamefont {G.}~\bibnamefont
  {Durastante}}, \bibinfo {author} {\bibfnamefont {C.}~\bibnamefont {Politi}},
  \bibinfo {author} {\bibfnamefont {M.}~\bibnamefont {Sohmen}}, \bibinfo
  {author} {\bibfnamefont {M.~J.}\ \bibnamefont {Mark}}, \ and\ \bibinfo
  {author} {\bibfnamefont {F.}~\bibnamefont {Ferlaino}},\ }\href {\doibase %
  10.1103/PhysRevLett.121.213601} {\bibfield  {journal} {\bibinfo  {journal}
  {Phys. Rev. Lett.}\ }\textbf {\bibinfo {volume} {121}},\ \bibinfo {pages}
  {213601} (\bibinfo {year} {2018})}\BibitemShut {NoStop}%
\bibitem [{\citenamefont {Durastante}\ \emph {et~al.}(2020)\citenamefont
  {Durastante}, \citenamefont {Politi}, \citenamefont {Sohmen}, \citenamefont
  {Ilzh\"ofer}, \citenamefont {Mark}, \citenamefont {Norcia},\ and\
  \citenamefont {Ferlaino}}]{Durastante2020}%
  \BibitemOpen
  \bibfield  {author} {\bibinfo {author} {\bibfnamefont {G.}~\bibnamefont
  {Durastante}}, \bibinfo {author} {\bibfnamefont {C.}~\bibnamefont {Politi}},
  \bibinfo {author} {\bibfnamefont {M.}~\bibnamefont {Sohmen}}, \bibinfo
  {author} {\bibfnamefont {P.}~\bibnamefont {Ilzh\"ofer}}, \bibinfo {author}
  {\bibfnamefont {M.~J.}\ \bibnamefont {Mark}}, \bibinfo {author}
  {\bibfnamefont {M.~A.}\ \bibnamefont {Norcia}}, \ and\ \bibinfo {author}
  {\bibfnamefont {F.}~\bibnamefont {Ferlaino}},\ }\href {\doibase %
  10.1103/PhysRevA.102.033330} {\bibfield  {journal} {\bibinfo  {journal}
  {Phys. Rev. A}\ }\textbf {\bibinfo {volume} {102}},\ \bibinfo {pages}
  {033330} (\bibinfo {year} {2020})}\BibitemShut {NoStop}%
\bibitem [{\citenamefont {Kadau}\ \emph {et~al.}(2016)\citenamefont {Kadau},
  \citenamefont {Schmitt}, \citenamefont {Wenzel}, \citenamefont {Wink},
  \citenamefont {Maier}, \citenamefont {Ferrier-Barbut},\ and\ \citenamefont
  {Pfau}}]{Kadau2016a}%
  \BibitemOpen
  \bibfield  {author} {\bibinfo {author} {\bibfnamefont {H.}~\bibnamefont
  {Kadau}}, \bibinfo {author} {\bibfnamefont {M.}~\bibnamefont {Schmitt}},
  \bibinfo {author} {\bibfnamefont {M.}~\bibnamefont {Wenzel}}, \bibinfo
  {author} {\bibfnamefont {C.}~\bibnamefont {Wink}}, \bibinfo {author}
  {\bibfnamefont {T.}~\bibnamefont {Maier}}, \bibinfo {author} {\bibfnamefont
  {I.}~\bibnamefont {Ferrier-Barbut}}, \ and\ \bibinfo {author} {\bibfnamefont
  {T.}~\bibnamefont {Pfau}},\ }\href {http://dx.doi.org/10.1038/nature16485}
  {\bibfield  {journal} {\bibinfo  {journal} {Nature}\ }\textbf {\bibinfo
  {volume} {530}},\ \bibinfo {pages} {194} (\bibinfo {year}
  {2016})}\BibitemShut {NoStop}%
\bibitem [{\citenamefont {Chomaz}\ \emph {et~al.}(2016)\citenamefont {Chomaz},
  \citenamefont {Baier}, \citenamefont {Petter}, \citenamefont {Mark},
  \citenamefont {W\"achtler}, \citenamefont {Santos},\ and\ \citenamefont
  {Ferlaino}}]{Chomaz2016}%
  \BibitemOpen
  \bibfield  {author} {\bibinfo {author} {\bibfnamefont {L.}~\bibnamefont
  {Chomaz}}, \bibinfo {author} {\bibfnamefont {S.}~\bibnamefont {Baier}},
  \bibinfo {author} {\bibfnamefont {D.}~\bibnamefont {Petter}}, \bibinfo
  {author} {\bibfnamefont {M.~J.}\ \bibnamefont {Mark}}, \bibinfo {author}
  {\bibfnamefont {F.}~\bibnamefont {W\"achtler}}, \bibinfo {author}
  {\bibfnamefont {L.}~\bibnamefont {Santos}}, \ and\ \bibinfo {author}
  {\bibfnamefont {F.}~\bibnamefont {Ferlaino}},\ }\href {\doibase %
  10.1103/PhysRevX.6.041039} {\bibfield  {journal} {\bibinfo  {journal} {Phys.
  Rev. X}\ }\textbf {\bibinfo {volume} {6}},\ \bibinfo {pages} {041039}
  (\bibinfo {year} {2016})}\BibitemShut {NoStop}%
\bibitem [{\citenamefont {Schmitt}\ \emph {et~al.}(2016)\citenamefont
  {Schmitt}, \citenamefont {Wenzel}, \citenamefont {B{\"o}ttcher},
  \citenamefont {Ferrier-Barbut},\ and\ \citenamefont {Pfau}}]{Schmitt2016a}%
  \BibitemOpen
  \bibfield  {author} {\bibinfo {author} {\bibfnamefont {M.}~\bibnamefont
  {Schmitt}}, \bibinfo {author} {\bibfnamefont {M.}~\bibnamefont {Wenzel}},
  \bibinfo {author} {\bibfnamefont {F.}~\bibnamefont {B{\"o}ttcher}}, \bibinfo
  {author} {\bibfnamefont {I.}~\bibnamefont {Ferrier-Barbut}}, \ and\ \bibinfo
  {author} {\bibfnamefont {T.}~\bibnamefont {Pfau}},\ }\href
  {http://dx.doi.org/10.1038/nature20126} {\bibfield  {journal} {\bibinfo
  {journal} {Nature}\ }\textbf {\bibinfo {volume} {539}},\ \bibinfo {pages}
  {259} (\bibinfo {year} {2016})}\BibitemShut {NoStop}%
\bibitem [{\citenamefont {Chomaz}\ \emph {et~al.}(2018)\citenamefont {Chomaz},
  \citenamefont {van Bijnen}, \citenamefont {Petter}, \citenamefont {Faraoni},
  \citenamefont {Baier}, \citenamefont {Becher}, \citenamefont {Mark},
  \citenamefont {W{\"a}chtler}, \citenamefont {Santos},\ and\ \citenamefont
  {Ferlaino}}]{Chomaz2018a}%
  \BibitemOpen
  \bibfield  {author} {\bibinfo {author} {\bibfnamefont {L.}~\bibnamefont
  {Chomaz}}, \bibinfo {author} {\bibfnamefont {R.~M.~W.}\ \bibnamefont {van
  Bijnen}}, \bibinfo {author} {\bibfnamefont {D.}~\bibnamefont {Petter}},
  \bibinfo {author} {\bibfnamefont {G.}~\bibnamefont {Faraoni}}, \bibinfo
  {author} {\bibfnamefont {S.}~\bibnamefont {Baier}}, \bibinfo {author}
  {\bibfnamefont {J.~H.}\ \bibnamefont {Becher}}, \bibinfo {author}
  {\bibfnamefont {M.~J.}\ \bibnamefont {Mark}}, \bibinfo {author}
  {\bibfnamefont {F.}~\bibnamefont {W{\"a}chtler}}, \bibinfo {author}
  {\bibfnamefont {L.}~\bibnamefont {Santos}}, \ and\ \bibinfo {author}
  {\bibfnamefont {F.}~\bibnamefont {Ferlaino}},\ }\href {\doibase %
  10.1038/s41567-018-0054-7} {\bibfield  {journal} {\bibinfo  {journal} {Nature
  Physics}\ }\textbf {\bibinfo {volume} {14}},\ \bibinfo {pages} {442}
  (\bibinfo {year} {2018})}\BibitemShut {NoStop}%
\bibitem [{\citenamefont {Petter}\ \emph {et~al.}(2019)\citenamefont {Petter},
  \citenamefont {Natale}, \citenamefont {van Bijnen}, \citenamefont
  {Patscheider}, \citenamefont {Mark}, \citenamefont {Chomaz},\ and\
  \citenamefont {Ferlaino}}]{Petter2019}%
  \BibitemOpen
  \bibfield  {author} {\bibinfo {author} {\bibfnamefont {D.}~\bibnamefont
  {Petter}}, \bibinfo {author} {\bibfnamefont {G.}~\bibnamefont {Natale}},
  \bibinfo {author} {\bibfnamefont {R.~M.~W.}\ \bibnamefont {van Bijnen}},
  \bibinfo {author} {\bibfnamefont {A.}~\bibnamefont {Patscheider}}, \bibinfo
  {author} {\bibfnamefont {M.~J.}\ \bibnamefont {Mark}}, \bibinfo {author}
  {\bibfnamefont {L.}~\bibnamefont {Chomaz}}, \ and\ \bibinfo {author}
  {\bibfnamefont {F.}~\bibnamefont {Ferlaino}},\ }\href {\doibase %
  10.1103/PhysRevLett.122.183401} {\bibfield  {journal} {\bibinfo  {journal}
  {Phys. Rev. Lett.}\ }\textbf {\bibinfo {volume} {122}},\ \bibinfo {pages}
  {183401} (\bibinfo {year} {2019})}\BibitemShut {NoStop}%
\bibitem [{\citenamefont {Tanzi}\ \emph {et~al.}(2019)\citenamefont {Tanzi},
  \citenamefont {Lucioni}, \citenamefont {Fam\`a}, \citenamefont {Catani},
  \citenamefont {Fioretti}, \citenamefont {Gabbanini}, \citenamefont {Bisset},
  \citenamefont {Santos},\ and\ \citenamefont {Modugno}}]{Tanzi2019}%
  \BibitemOpen
  \bibfield  {author} {\bibinfo {author} {\bibfnamefont {L.}~\bibnamefont
  {Tanzi}}, \bibinfo {author} {\bibfnamefont {E.}~\bibnamefont {Lucioni}},
  \bibinfo {author} {\bibfnamefont {F.}~\bibnamefont {Fam\`a}}, \bibinfo
  {author} {\bibfnamefont {J.}~\bibnamefont {Catani}}, \bibinfo {author}
  {\bibfnamefont {A.}~\bibnamefont {Fioretti}}, \bibinfo {author}
  {\bibfnamefont {C.}~\bibnamefont {Gabbanini}}, \bibinfo {author}
  {\bibfnamefont {R.~N.}\ \bibnamefont {Bisset}}, \bibinfo {author}
  {\bibfnamefont {L.}~\bibnamefont {Santos}}, \ and\ \bibinfo {author}
  {\bibfnamefont {G.}~\bibnamefont {Modugno}},\ }\href {\doibase %
  10.1103/PhysRevLett.122.130405} {\bibfield  {journal} {\bibinfo  {journal}
  {Phys. Rev. Lett.}\ }\textbf {\bibinfo {volume} {122}},\ \bibinfo {pages}
  {130405} (\bibinfo {year} {2019})}\BibitemShut {NoStop}%
\bibitem [{\citenamefont {B\"ottcher}\ \emph {et~al.}(2019)\citenamefont
  {B\"ottcher}, \citenamefont {Schmidt}, \citenamefont {Wenzel}, \citenamefont
  {Hertkorn}, \citenamefont {Guo}, \citenamefont {Langen},\ and\ \citenamefont
  {Pfau}}]{Bottcher2019}%
  \BibitemOpen
  \bibfield  {author} {\bibinfo {author} {\bibfnamefont {F.}~\bibnamefont
  {B\"ottcher}}, \bibinfo {author} {\bibfnamefont {J.-N.}\ \bibnamefont
  {Schmidt}}, \bibinfo {author} {\bibfnamefont {M.}~\bibnamefont {Wenzel}},
  \bibinfo {author} {\bibfnamefont {J.}~\bibnamefont {Hertkorn}}, \bibinfo
  {author} {\bibfnamefont {M.}~\bibnamefont {Guo}}, \bibinfo {author}
  {\bibfnamefont {T.}~\bibnamefont {Langen}}, \ and\ \bibinfo {author}
  {\bibfnamefont {T.}~\bibnamefont {Pfau}},\ }\href {\doibase %
  10.1103/PhysRevX.9.011051} {\bibfield  {journal} {\bibinfo  {journal} {Phys.
  Rev. X}\ }\textbf {\bibinfo {volume} {9}},\ \bibinfo {pages} {011051}
  (\bibinfo {year} {2019})}\BibitemShut {NoStop}%
\bibitem [{\citenamefont {Chomaz}\ \emph {et~al.}(2019)\citenamefont {Chomaz},
  \citenamefont {Petter}, \citenamefont {Ilzh\"ofer}, \citenamefont {Natale},
  \citenamefont {Trautmann}, \citenamefont {Politi}, \citenamefont
  {Durastante}, \citenamefont {van Bijnen}, \citenamefont {Patscheider},
  \citenamefont {Sohmen}, \citenamefont {Mark},\ and\ \citenamefont
  {Ferlaino}}]{Chomaz2019}%
  \BibitemOpen
  \bibfield  {author} {\bibinfo {author} {\bibfnamefont {L.}~\bibnamefont
  {Chomaz}}, \bibinfo {author} {\bibfnamefont {D.}~\bibnamefont {Petter}},
  \bibinfo {author} {\bibfnamefont {P.}~\bibnamefont {Ilzh\"ofer}}, \bibinfo
  {author} {\bibfnamefont {G.}~\bibnamefont {Natale}}, \bibinfo {author}
  {\bibfnamefont {A.}~\bibnamefont {Trautmann}}, \bibinfo {author}
  {\bibfnamefont {C.}~\bibnamefont {Politi}}, \bibinfo {author} {\bibfnamefont
  {G.}~\bibnamefont {Durastante}}, \bibinfo {author} {\bibfnamefont {R.~M.~W.}\
  \bibnamefont {van Bijnen}}, \bibinfo {author} {\bibfnamefont
  {A.}~\bibnamefont {Patscheider}}, \bibinfo {author} {\bibfnamefont
  {M.}~\bibnamefont {Sohmen}}, \bibinfo {author} {\bibfnamefont {M.~J.}\
  \bibnamefont {Mark}}, \ and\ \bibinfo {author} {\bibfnamefont
  {F.}~\bibnamefont {Ferlaino}},\ }\href {\doibase 10.1103/PhysRevX.9.021012}
  {\bibfield  {journal} {\bibinfo  {journal} {Phys. Rev. X}\ }\textbf {\bibinfo
  {volume} {9}},\ \bibinfo {pages} {021012} (\bibinfo {year}
  {2019})}\BibitemShut {NoStop}%
\bibitem [{\citenamefont {G\'oral}\ and\ \citenamefont
  {Santos}(2002)}]{Goral2002}%
  \BibitemOpen
  \bibfield  {author} {\bibinfo {author} {\bibfnamefont {K.}~\bibnamefont
  {G\'oral}}\ and\ \bibinfo {author} {\bibfnamefont {L.}~\bibnamefont
  {Santos}},\ }\href {\doibase 10.1103/PhysRevA.66.023613} {\bibfield
  {journal} {\bibinfo  {journal} {Phys. Rev. A}\ }\textbf {\bibinfo {volume}
  {66}},\ \bibinfo {pages} {023613} (\bibinfo {year} {2002})}\BibitemShut
  {NoStop}%
\bibitem [{\citenamefont {Wilson}\ \emph {et~al.}(2012)\citenamefont {Wilson},
  \citenamefont {Ticknor}, \citenamefont {Bohn},\ and\ \citenamefont
  {Timmermans}}]{Wilson2012a}%
  \BibitemOpen
  \bibfield  {author} {\bibinfo {author} {\bibfnamefont {R.~M.}\ \bibnamefont
  {Wilson}}, \bibinfo {author} {\bibfnamefont {C.}~\bibnamefont {Ticknor}},
  \bibinfo {author} {\bibfnamefont {J.~L.}\ \bibnamefont {Bohn}}, \ and\
  \bibinfo {author} {\bibfnamefont {E.}~\bibnamefont {Timmermans}},\ }\href
  {\doibase 10.1103/PhysRevA.86.033606} {\bibfield  {journal} {\bibinfo
  {journal} {Phys. Rev. A}\ }\textbf {\bibinfo {volume} {86}},\ \bibinfo
  {pages} {033606} (\bibinfo {year} {2012})}\BibitemShut {NoStop}%
\bibitem [{\citenamefont {Saito}\ \emph {et~al.}(2009)\citenamefont {Saito},
  \citenamefont {Kawaguchi},\ and\ \citenamefont {Ueda}}]{Saito2009}%
  \BibitemOpen
  \bibfield  {author} {\bibinfo {author} {\bibfnamefont {H.}~\bibnamefont
  {Saito}}, \bibinfo {author} {\bibfnamefont {Y.}~\bibnamefont {Kawaguchi}}, \
  and\ \bibinfo {author} {\bibfnamefont {M.}~\bibnamefont {Ueda}},\ }\href
  {\doibase 10.1103/PhysRevLett.102.230403} {\bibfield  {journal} {\bibinfo
  {journal} {Phys. Rev. Lett.}\ }\textbf {\bibinfo {volume} {102}},\ \bibinfo
  {pages} {230403} (\bibinfo {year} {2009})}\BibitemShut {NoStop}%
\bibitem [{\citenamefont {Xi}\ \emph {et~al.}(2018)\citenamefont {Xi},
  \citenamefont {Byrnes},\ and\ \citenamefont {Saito}}]{KuiTian2018}%
  \BibitemOpen
  \bibfield  {author} {\bibinfo {author} {\bibfnamefont {K.-T.}\ \bibnamefont
  {Xi}}, \bibinfo {author} {\bibfnamefont {T.}~\bibnamefont {Byrnes}}, \ and\
  \bibinfo {author} {\bibfnamefont {H.}~\bibnamefont {Saito}},\ }\href
  {\doibase 10.1103/PhysRevA.97.023625} {\bibfield  {journal} {\bibinfo
  {journal} {Phys. Rev. A}\ }\textbf {\bibinfo {volume} {97}},\ \bibinfo
  {pages} {023625} (\bibinfo {year} {2018})}\BibitemShut {NoStop}%
\bibitem [{\citenamefont {Zhang}\ \emph {et~al.}(2015)\citenamefont {Zhang},
  \citenamefont {Han}, \citenamefont {Wen}, \citenamefont {Zhang},
  \citenamefont {Dong}, \citenamefont {Chang},\ and\ \citenamefont
  {Zhang}}]{Zhang2015}%
  \BibitemOpen
  \bibfield  {author} {\bibinfo {author} {\bibfnamefont {X.-F.}\ \bibnamefont
  {Zhang}}, \bibinfo {author} {\bibfnamefont {W.}~\bibnamefont {Han}}, \bibinfo
  {author} {\bibfnamefont {L.}~\bibnamefont {Wen}}, \bibinfo {author}
  {\bibfnamefont {P.}~\bibnamefont {Zhang}}, \bibinfo {author} {\bibfnamefont
  {R.-F.}\ \bibnamefont {Dong}}, \bibinfo {author} {\bibfnamefont
  {H.}~\bibnamefont {Chang}}, \ and\ \bibinfo {author} {\bibfnamefont {S.-G.}\
  \bibnamefont {Zhang}},\ }\href {\doibase 10.1038/srep08684} {\bibfield
  {journal} {\bibinfo  {journal} {Scientific Reports}\ }\textbf {\bibinfo
  {volume} {5}},\ \bibinfo {pages} {8684} (\bibinfo {year} {2015})}\BibitemShut
  {NoStop}%
\bibitem [{\citenamefont {Zhang}\ \emph {et~al.}(2016)\citenamefont {Zhang},
  \citenamefont {Wen}, \citenamefont {Dai}, \citenamefont {Dong}, \citenamefont
  {Jiang}, \citenamefont {Chang},\ and\ \citenamefont {Zhang}}]{Zhang2016}%
  \BibitemOpen
  \bibfield  {author} {\bibinfo {author} {\bibfnamefont {X.-F.}\ \bibnamefont
  {Zhang}}, \bibinfo {author} {\bibfnamefont {L.}~\bibnamefont {Wen}}, \bibinfo
  {author} {\bibfnamefont {C.-Q.}\ \bibnamefont {Dai}}, \bibinfo {author}
  {\bibfnamefont {R.-F.}\ \bibnamefont {Dong}}, \bibinfo {author}
  {\bibfnamefont {H.-F.}\ \bibnamefont {Jiang}}, \bibinfo {author}
  {\bibfnamefont {H.}~\bibnamefont {Chang}}, \ and\ \bibinfo {author}
  {\bibfnamefont {S.-G.}\ \bibnamefont {Zhang}},\ }\href {\doibase %
  10.1038/srep19380} {\bibfield  {journal} {\bibinfo  {journal} {Scientific
  Reports}\ }\textbf {\bibinfo {volume} {6}},\ \bibinfo {pages} {19380}
  (\bibinfo {year} {2016})}\BibitemShut {NoStop}%
\bibitem [{\citenamefont {Kumar}\ \emph
  {et~al.}(2017{\natexlab{a}})\citenamefont {Kumar}, \citenamefont {Tomio},
  \citenamefont {Malomed},\ and\ \citenamefont {Gammal}}]{Kumar2017}%
  \BibitemOpen
  \bibfield  {author} {\bibinfo {author} {\bibfnamefont {R.~K.}\ \bibnamefont
  {Kumar}}, \bibinfo {author} {\bibfnamefont {L.}~\bibnamefont {Tomio}},
  \bibinfo {author} {\bibfnamefont {B.~A.}\ \bibnamefont {Malomed}}, \ and\
  \bibinfo {author} {\bibfnamefont {A.}~\bibnamefont {Gammal}},\ }\href
  {\doibase 10.1103/PhysRevA.96.063624} {\bibfield  {journal} {\bibinfo
  {journal} {Phys. Rev. A}\ }\textbf {\bibinfo {volume} {96}},\ \bibinfo
  {pages} {063624} (\bibinfo {year} {2017}{\natexlab{a}})}\BibitemShut
  {NoStop}%
\bibitem [{\citenamefont {Shirley}\ \emph {et~al.}(2014)\citenamefont
  {Shirley}, \citenamefont {Anderson}, \citenamefont {Clark},\ and\
  \citenamefont {Wilson}}]{Shirley2014}%
  \BibitemOpen
  \bibfield  {author} {\bibinfo {author} {\bibfnamefont {W.~E.}\ \bibnamefont
  {Shirley}}, \bibinfo {author} {\bibfnamefont {B.~M.}\ \bibnamefont
  {Anderson}}, \bibinfo {author} {\bibfnamefont {C.~W.}\ \bibnamefont {Clark}},
  \ and\ \bibinfo {author} {\bibfnamefont {R.~M.}\ \bibnamefont {Wilson}},\
  }\href {\doibase 10.1103/PhysRevLett.113.165301} {\bibfield  {journal}
  {\bibinfo  {journal} {Phys. Rev. Lett.}\ }\textbf {\bibinfo {volume} {113}},\
  \bibinfo {pages} {165301} (\bibinfo {year} {2014})}\BibitemShut {NoStop}%
\bibitem [{\citenamefont {Tomio}\ \emph {et~al.}(2020)\citenamefont {Tomio},
  \citenamefont {Kumar},\ and\ \citenamefont {Gammal}}]{Tomio2020}%
  \BibitemOpen
  \bibfield  {author} {\bibinfo {author} {\bibfnamefont {L.}~\bibnamefont
  {Tomio}}, \bibinfo {author} {\bibfnamefont {R.~K.}\ \bibnamefont {Kumar}}, \
  and\ \bibinfo {author} {\bibfnamefont {A.}~\bibnamefont {Gammal}},\ }\href
  {\doibase 10.21468/SciPostPhysProc.3.023} {\bibfield  {journal} {\bibinfo
  {journal} {SciPost Phys. Proc.}\ ,\ \bibinfo {pages} {23}} (\bibinfo {year}
  {2020})}\BibitemShut {NoStop}%
\bibitem [{\citenamefont {Kumar}\ \emph {et~al.}(2019)\citenamefont {Kumar},
  \citenamefont {Tomio},\ and\ \citenamefont {Gammal}}]{Kumar2019}%
  \BibitemOpen
  \bibfield  {author} {\bibinfo {author} {\bibfnamefont {R.~K.}\ \bibnamefont
  {Kumar}}, \bibinfo {author} {\bibfnamefont {L.}~\bibnamefont {Tomio}}, \ and\
  \bibinfo {author} {\bibfnamefont {A.}~\bibnamefont {Gammal}},\ }\href
  {\doibase 10.1103/PhysRevA.99.043606} {\bibfield  {journal} {\bibinfo
  {journal} {Phys. Rev. A}\ }\textbf {\bibinfo {volume} {99}},\ \bibinfo
  {pages} {043606} (\bibinfo {year} {2019})}\BibitemShut {NoStop}%
\bibitem [{\citenamefont {Giovanazzi}\ \emph {et~al.}(2002)\citenamefont
  {Giovanazzi}, \citenamefont {G\"orlitz},\ and\ \citenamefont
  {Pfau}}]{Giovanazzi2002}%
  \BibitemOpen
  \bibfield  {author} {\bibinfo {author} {\bibfnamefont {S.}~\bibnamefont
  {Giovanazzi}}, \bibinfo {author} {\bibfnamefont {A.}~\bibnamefont
  {G\"orlitz}}, \ and\ \bibinfo {author} {\bibfnamefont {T.}~\bibnamefont
  {Pfau}},\ }\href {\doibase 10.1103/PhysRevLett.89.130401} {\bibfield
  {journal} {\bibinfo  {journal} {Phys. Rev. Lett.}\ }\textbf {\bibinfo
  {volume} {89}},\ \bibinfo {pages} {130401} (\bibinfo {year}
  {2002})}\BibitemShut {NoStop}%
\bibitem [{\citenamefont {Tang}\ \emph {et~al.}(2018)\citenamefont {Tang},
  \citenamefont {Kao}, \citenamefont {Li},\ and\ \citenamefont
  {Lev}}]{Tang2018}%
  \BibitemOpen
  \bibfield  {author} {\bibinfo {author} {\bibfnamefont {Y.}~\bibnamefont
  {Tang}}, \bibinfo {author} {\bibfnamefont {W.}~\bibnamefont {Kao}}, \bibinfo
  {author} {\bibfnamefont {K.-Y.}\ \bibnamefont {Li}}, \ and\ \bibinfo {author}
  {\bibfnamefont {B.~L.}\ \bibnamefont {Lev}},\ }\href {\doibase %
  10.1103/PhysRevLett.120.230401} {\bibfield  {journal} {\bibinfo  {journal}
  {Phys. Rev. Lett.}\ }\textbf {\bibinfo {volume} {120}},\ \bibinfo {pages}
  {230401} (\bibinfo {year} {2018})}\BibitemShut {NoStop}%
\bibitem [{\citenamefont {Jain}\ and\ \citenamefont
  {Boninsegni}(2011)}]{Jain2011}%
  \BibitemOpen
  \bibfield  {author} {\bibinfo {author} {\bibfnamefont {P.}~\bibnamefont
  {Jain}}\ and\ \bibinfo {author} {\bibfnamefont {M.}~\bibnamefont
  {Boninsegni}},\ }\href {\doibase 10.1103/PhysRevA.83.023602} {\bibfield
  {journal} {\bibinfo  {journal} {Phys. Rev. A}\ }\textbf {\bibinfo {volume}
  {83}},\ \bibinfo {pages} {023602} (\bibinfo {year} {2011})}\BibitemShut
  {NoStop}%
\bibitem [{\citenamefont {Young-S.}\ and\ \citenamefont
  {Adhikari}(2012)}]{Young-S2012}%
  \BibitemOpen
  \bibfield  {author} {\bibinfo {author} {\bibfnamefont {L.~E.}\ \bibnamefont
  {Young-S.}}\ and\ \bibinfo {author} {\bibfnamefont {S.~K.}\ \bibnamefont
  {Adhikari}},\ }\href {\doibase 10.1103/PhysRevA.86.063611} {\bibfield
  {journal} {\bibinfo  {journal} {Phys. Rev. A}\ }\textbf {\bibinfo {volume}
  {86}},\ \bibinfo {pages} {063611} (\bibinfo {year} {2012})}\BibitemShut
  {NoStop}%
\bibitem [{\citenamefont {Kumar}\ \emph
  {et~al.}(2017{\natexlab{b}})\citenamefont {Kumar}, \citenamefont
  {Muruganandam}, \citenamefont {Tomio},\ and\ \citenamefont
  {Gammal}}]{Kumar2017B}%
  \BibitemOpen
  \bibfield  {author} {\bibinfo {author} {\bibfnamefont {R.~K.}\ \bibnamefont
  {Kumar}}, \bibinfo {author} {\bibfnamefont {P.}~\bibnamefont {Muruganandam}},
  \bibinfo {author} {\bibfnamefont {L.}~\bibnamefont {Tomio}}, \ and\ \bibinfo
  {author} {\bibfnamefont {A.}~\bibnamefont {Gammal}},\ }\href {\doibase %
  10.1088/2399-6528/aa8db5} {\bibfield  {journal} {\bibinfo  {journal} {J.
  Phys. Commun.}\ }\textbf {\bibinfo {volume} {1}},\ \bibinfo {pages} {035012}
  (\bibinfo {year} {2017}{\natexlab{b}})}\BibitemShut {NoStop}%
\bibitem [{\citenamefont {Gligori\ifmmode~\acute{c}\else \'{c}\fi{}}\ \emph
  {et~al.}(2010)\citenamefont {Gligori\ifmmode~\acute{c}\else \'{c}\fi{}},
  \citenamefont {Maluckov}, \citenamefont {Stepi\ifmmode~\acute{c}\else
  \'{c}\fi{}}, \citenamefont {Had\ifmmode~\check{z}\else \v{z}\fi{}ievski},\
  and\ \citenamefont {Malomed}}]{Gligoric2010}%
  \BibitemOpen
  \bibfield  {author} {\bibinfo {author} {\bibfnamefont {G.}~\bibnamefont
  {Gligori\ifmmode~\acute{c}\else \'{c}\fi{}}}, \bibinfo {author}
  {\bibfnamefont {A.}~\bibnamefont {Maluckov}}, \bibinfo {author}
  {\bibfnamefont {M.}~\bibnamefont {Stepi\ifmmode~\acute{c}\else \'{c}\fi{}}},
  \bibinfo {author} {\bibfnamefont {L.~c.~v.}\ \bibnamefont
  {Had\ifmmode~\check{z}\else \v{z}\fi{}ievski}}, \ and\ \bibinfo {author}
  {\bibfnamefont {B.~A.}\ \bibnamefont {Malomed}},\ }\href {\doibase %
  10.1103/PhysRevA.82.033624} {\bibfield  {journal} {\bibinfo  {journal} {Phys.
  Rev. A}\ }\textbf {\bibinfo {volume} {82}},\ \bibinfo {pages} {033624}
  (\bibinfo {year} {2010})}\BibitemShut {NoStop}%
\bibitem [{\citenamefont {Hocine}\ and\ \citenamefont
  {Benarous}(2019)}]{Hocine2019}%
  \BibitemOpen
  \bibfield  {author} {\bibinfo {author} {\bibfnamefont {A.}~\bibnamefont
  {Hocine}}\ and\ \bibinfo {author} {\bibfnamefont {M.}~\bibnamefont
  {Benarous}},\ }\href {\doibase 10.1007/s10909-018-2088-8} {\bibfield
  {journal} {\bibinfo  {journal} {Journal of Low Temperature Physics}\ }\textbf
  {\bibinfo {volume} {194}},\ \bibinfo {pages} {209} (\bibinfo {year}
  {2019})}\BibitemShut {NoStop}%
\bibitem [{\citenamefont {Chalopin}\ \emph {et~al.}(2020)\citenamefont
  {Chalopin}, \citenamefont {Satoor}, \citenamefont {Evrard}, \citenamefont
  {Makhalov}, \citenamefont {Dalibard}, \citenamefont {Lopes},\ and\
  \citenamefont {Nascimbene}}]{Chalopin2020}%
  \BibitemOpen
  \bibfield  {author} {\bibinfo {author} {\bibfnamefont {T.}~\bibnamefont
  {Chalopin}}, \bibinfo {author} {\bibfnamefont {T.}~\bibnamefont {Satoor}},
  \bibinfo {author} {\bibfnamefont {A.}~\bibnamefont {Evrard}}, \bibinfo
  {author} {\bibfnamefont {V.}~\bibnamefont {Makhalov}}, \bibinfo {author}
  {\bibfnamefont {J.}~\bibnamefont {Dalibard}}, \bibinfo {author}
  {\bibfnamefont {R.}~\bibnamefont {Lopes}}, \ and\ \bibinfo {author}
  {\bibfnamefont {S.}~\bibnamefont {Nascimbene}},\ }\href {\doibase %
  10.1038/s41567-020-0942-5} {\bibfield  {journal} {\bibinfo  {journal} {Nature
  Physics}\ }\textbf {\bibinfo {volume} {16}},\ \bibinfo {pages} {1017}
  (\bibinfo {year} {2020})}\BibitemShut {NoStop}%
\bibitem [{\citenamefont {Grimm}\ \emph {et~al.}(2000)\citenamefont {Grimm},
  \citenamefont {Weidem{\"u}ller},\ and\ \citenamefont
  {Ovchinnikov}}]{Grimm2000}%
  \BibitemOpen
  \bibfield  {author} {\bibinfo {author} {\bibfnamefont {R.}~\bibnamefont
  {Grimm}}, \bibinfo {author} {\bibfnamefont {M.}~\bibnamefont
  {Weidem{\"u}ller}}, \ and\ \bibinfo {author} {\bibfnamefont {Y.~B.}\
  \bibnamefont {Ovchinnikov}},\ }\href {\doibase %
  https://doi.org/10.1016/S1049-250X(08)60186-X} {\bibfield  {journal}
  {\bibinfo  {journal} {Adv. At. Mol. Opt. Phys.}\ }\textbf {\bibinfo {volume}
  {42}},\ \bibinfo {pages} {95 } (\bibinfo {year} {2000})}\BibitemShut
  {NoStop}%
\bibitem [{\citenamefont {Bao}\ \emph {et~al.}(2006)\citenamefont {Bao},
  \citenamefont {Chern},\ and\ \citenamefont {Lim}}]{Bao2006}%
  \BibitemOpen
  \bibfield  {author} {\bibinfo {author} {\bibfnamefont {W.}~\bibnamefont
  {Bao}}, \bibinfo {author} {\bibfnamefont {I.-L.}\ \bibnamefont {Chern}}, \
  and\ \bibinfo {author} {\bibfnamefont {F.~Y.}\ \bibnamefont {Lim}},\ }\href
  {\doibase https://doi.org/10.1016/j.jcp.2006.04.019} {\bibfield  {journal}
  {\bibinfo  {journal} {J. Comput. Phys.}\ }\textbf {\bibinfo {volume} {219}},\
  \bibinfo {pages} {836 } (\bibinfo {year} {2006})}\BibitemShut {NoStop}%
\bibitem [{\citenamefont {Lee}\ \emph {et~al.}()\citenamefont {Lee},
  \citenamefont {Baillie},\ and\ \citenamefont {Blakie}}]{Lee2020}%
  \BibitemOpen
  \bibfield  {author} {\bibinfo {author} {\bibfnamefont {A.-C.}\ \bibnamefont
  {Lee}}, \bibinfo {author} {\bibfnamefont {D.}~\bibnamefont {Baillie}}, \ and\
  \bibinfo {author} {\bibfnamefont {P.~B.}\ \bibnamefont {Blakie}},\
  }\href@noop {} {\enquote {\bibinfo {title} {Numerical calculation of dipolar
  quantum droplet stationary states},}\ }\Eprint
  {http://arxiv.org/abs/2012.11103} {arXiv:2012.11103} \BibitemShut {NoStop}%
\bibitem [{\citenamefont {Kelley}(2003)}]{Kelley2003}%
  \BibitemOpen
  \bibfield  {author} {\bibinfo {author} {\bibfnamefont {C.~T.}\ \bibnamefont
  {Kelley}},\ }\href {\doibase 10.1137/1.9780898718898} {\emph {\bibinfo
  {title} {Solving Nonlinear Equations with Newton's Method}}}\ (\bibinfo
  {publisher} {SIAM},\ \bibinfo {year} {2003})\BibitemShut {NoStop}%
\bibitem [{\citenamefont {Ronen}\ \emph {et~al.}(2006)\citenamefont {Ronen},
  \citenamefont {Bortolotti},\ and\ \citenamefont {Bohn}}]{Ronen2006a}%
  \BibitemOpen
  \bibfield  {author} {\bibinfo {author} {\bibfnamefont {S.}~\bibnamefont
  {Ronen}}, \bibinfo {author} {\bibfnamefont {D.~C.~E.}\ \bibnamefont
  {Bortolotti}}, \ and\ \bibinfo {author} {\bibfnamefont {J.~L.}\ \bibnamefont
  {Bohn}},\ }\href {\doibase 10.1103/PhysRevA.74.013623} {\bibfield  {journal}
  {\bibinfo  {journal} {Phys. Rev. A}\ }\textbf {\bibinfo {volume} {74}},\
  \bibinfo {pages} {013623} (\bibinfo {year} {2006})}\BibitemShut {NoStop}%
\bibitem [{\citenamefont {Lu}\ \emph {et~al.}(2010)\citenamefont {Lu},
  \citenamefont {Lu}, \citenamefont {Zhang}, \citenamefont {Qiu}, \citenamefont
  {Pu},\ and\ \citenamefont {Yi}}]{Lu2010a}%
  \BibitemOpen
  \bibfield  {author} {\bibinfo {author} {\bibfnamefont {H.-Y.}\ \bibnamefont
  {Lu}}, \bibinfo {author} {\bibfnamefont {H.}~\bibnamefont {Lu}}, \bibinfo
  {author} {\bibfnamefont {J.-N.}\ \bibnamefont {Zhang}}, \bibinfo {author}
  {\bibfnamefont {R.-Z.}\ \bibnamefont {Qiu}}, \bibinfo {author} {\bibfnamefont
  {H.}~\bibnamefont {Pu}}, \ and\ \bibinfo {author} {\bibfnamefont
  {S.}~\bibnamefont {Yi}},\ }\href {\doibase 10.1103/PhysRevA.82.023622}
  {\bibfield  {journal} {\bibinfo  {journal} {Phys. Rev. A}\ }\textbf {\bibinfo
  {volume} {82}},\ \bibinfo {pages} {023622} (\bibinfo {year}
  {2010})}\BibitemShut {NoStop}%
\bibitem [{\citenamefont {Lee}\ \emph {et~al.}(2016)\citenamefont {Lee},
  \citenamefont {J\o{}rgensen}, \citenamefont {Liu}, \citenamefont {Wacker},
  \citenamefont {Arlt},\ and\ \citenamefont {Proukakis}}]{Lee2016}%
  \BibitemOpen
  \bibfield  {author} {\bibinfo {author} {\bibfnamefont {K.~L.}\ \bibnamefont
  {Lee}}, \bibinfo {author} {\bibfnamefont {N.~B.}\ \bibnamefont
  {J\o{}rgensen}}, \bibinfo {author} {\bibfnamefont {I.Kang}\ \bibnamefont
  {Liu}}, \bibinfo {author} {\bibfnamefont {L.}~\bibnamefont {Wacker}},
  \bibinfo {author} {\bibfnamefont {J.~J.}\ \bibnamefont {Arlt}}, \ and\
  \bibinfo {author} {\bibfnamefont {N.~P.}\ \bibnamefont {Proukakis}},\ }\href
  {\doibase 10.1103/PhysRevA.94.013602} {\bibfield  {journal} {\bibinfo
  {journal} {Phys. Rev. A}\ }\textbf {\bibinfo {volume} {94}},\ \bibinfo
  {pages} {013602} (\bibinfo {year} {2016})}\BibitemShut {NoStop}%
\bibitem [{\citenamefont {Bisset}\ \emph {et~al.}(2018)\citenamefont {Bisset},
  \citenamefont {Kevrekidis},\ and\ \citenamefont {Ticknor}}]{Bisset2018}%
  \BibitemOpen
  \bibfield  {author} {\bibinfo {author} {\bibfnamefont {R.~N.}\ \bibnamefont
  {Bisset}}, \bibinfo {author} {\bibfnamefont {P.~G.}\ \bibnamefont
  {Kevrekidis}}, \ and\ \bibinfo {author} {\bibfnamefont {C.}~\bibnamefont
  {Ticknor}},\ }\href {\doibase 10.1103/PhysRevA.97.023602} {\bibfield
  {journal} {\bibinfo  {journal} {Phys. Rev. A}\ }\textbf {\bibinfo {volume}
  {97}},\ \bibinfo {pages} {023602} (\bibinfo {year} {2018})}\BibitemShut
  {NoStop}%
\bibitem [{\citenamefont {Santos}\ \emph {et~al.}(2003)\citenamefont {Santos},
  \citenamefont {Shlyapnikov},\ and\ \citenamefont {Lewenstein}}]{Santos2003a}%
  \BibitemOpen
  \bibfield  {author} {\bibinfo {author} {\bibfnamefont {L.}~\bibnamefont
  {Santos}}, \bibinfo {author} {\bibfnamefont {G.~V.}\ \bibnamefont
  {Shlyapnikov}}, \ and\ \bibinfo {author} {\bibfnamefont {M.}~\bibnamefont
  {Lewenstein}},\ }\href {\doibase 10.1103/PhysRevLett.90.250403} {\bibfield
  {journal} {\bibinfo  {journal} {Phys. Rev. Lett.}\ }\textbf {\bibinfo
  {volume} {90}},\ \bibinfo {pages} {250403} (\bibinfo {year}
  {2003})}\BibitemShut {NoStop}%
\bibitem [{\citenamefont {O'Dell}\ \emph {et~al.}(2003)\citenamefont
  {O\char39{}Dell}, \citenamefont {Giovanazzi},\ and\ \citenamefont
  {Kurizki}}]{ODell2003a}%
  \BibitemOpen
  \bibfield  {author} {\bibinfo {author} {\bibfnamefont {D.~H.~J.}\
  \bibnamefont {O'Dell}}, \bibinfo {author} {\bibfnamefont
  {S.}~\bibnamefont {Giovanazzi}}, \ and\ \bibinfo {author} {\bibfnamefont
  {G.}~\bibnamefont {Kurizki}},\ }\href {\doibase %
  10.1103/PhysRevLett.90.110402} {\bibfield  {journal} {\bibinfo  {journal}
  {Phys. Rev. Lett.}\ }\textbf {\bibinfo {volume} {90}},\ \bibinfo {pages}
  {110402} (\bibinfo {year} {2003})}\BibitemShut {NoStop}%
\bibitem [{\citenamefont {Bisset}\ \emph {et~al.}(2021)\citenamefont {Bisset},
  \citenamefont {Ardila},\ and\ \citenamefont {Santos}}]{Bisset2021}%
  \BibitemOpen
  \bibfield  {author} {\bibinfo {author} {\bibfnamefont {R.~N.}\ \bibnamefont
  {Bisset}}, \bibinfo {author} {\bibfnamefont {L.~A.~Pe{\~n}a}\ \bibnamefont
  {Ardila}}, \ and\ \bibinfo {author} {\bibfnamefont {L.}~\bibnamefont
  {Santos}},\ }\href {\doibase 10.1103/PhysRevLett.126.025301} {\bibfield
  {journal} {\bibinfo  {journal} {Phys. Rev. Lett.}\ }\textbf {\bibinfo
  {volume} {126}},\ \bibinfo {pages} {025301} (\bibinfo {year}
  {2021})}\BibitemShut {NoStop}%
\bibitem [{\citenamefont {Smith}\ \emph {et~al.}(2021)\citenamefont {Smith},
  \citenamefont {Baillie},\ and\ \citenamefont {Blakie}}]{Smith2021}%
  \BibitemOpen
  \bibfield  {author} {\bibinfo {author} {\bibfnamefont {J.~C.}\ \bibnamefont
  {Smith}}, \bibinfo {author} {\bibfnamefont {D.}~\bibnamefont {Baillie}}, \
  and\ \bibinfo {author} {\bibfnamefont {P.~B.}\ \bibnamefont {Blakie}},\
  }\href {\doibase 10.1103/PhysRevLett.126.025302} {\bibfield  {journal}
  {\bibinfo  {journal} {Phys. Rev. Lett.}\ }\textbf {\bibinfo {volume} {126}},\
  \bibinfo {pages} {025302} (\bibinfo {year} {2021})}\BibitemShut {NoStop}%
\end{thebibliography}

%

\end{document}